# STATISTICAL ANALYSIS OF AN ARCHEOLOGICAL FIND[1]

By Andrey Feuerverger

*University of Toronto*

In 1980, a burial tomb was unearthed in Jerusalem containing ossuaries (limestone coffins) bearing such inscriptions as Yeshua son of Yehosef, Marya, Yoseh—names which match those of New Testament (NT) figures, but were otherwise in common use. This paper discusses certain statistical aspects of authenticating or repudiating links between this find and the NT family. The available data are laid out, and we examine the distribution of names (onomasticon) of the era. An approach is proposed for measuring the "surprisingness" of the observed outcome relative to a "hypothesis" that the tombsite belonged to the NT family. On the basis of a particular—but far from uncontested—set of assumptions, our measure of "surprisingness" is significantly high.

**1. Introduction and summary.** In March 1980, the Solel Boneh Construction Company interrupted excavation work at an apartment site complex in the East Talpiyot neighbourhood of Jerusalem, and reported to Israel's Department of Antiquities and Museums that it had accidentally unearthed a previously unknown entrance to a burial cave. This tomb is located approximately 2.5 kilometers south of the site of the Second Temple in the Old City of Jerusalem, destroyed by the Romans in 70 CE.[2]

Shortly after its discovery, this burial site was examined and surveyed and salvage excavations were carried out. Within this cave a number of ossuaries[3] were found, some bearing inscriptions, and these were published

[2]CE and BCE are abbreviations for "common era" and "before the common era"—secular versions of the abbreviations AD and BC.

[3]Ossuaries are repositories for bones; see Section 3.

by Rahmani (1994), pages 222–224, Nos. 701–709 and by Kloner (1996).
Among these ossuaries were found such inscriptions as "Marya," "Yoseh,"
"Yeshua son of Yehosef," and other inscriptions of related interest.

Since the practice of ossuary burial was prevalent among Jews at the time
Jesus of Nazareth was crucified in Jerusalem at the behest of the Romans,
archeological questions arise in respect of the identity of the individuals
buried in this tomb. Since names such as Yehosef, Marya, Yeshua, etc.,
were not uncommon during the era in which such burials took place, the
task of assessing whether or not these ossuaries might be those of the New
Testament (NT) family is not straightforward.

Several disciplines bear on assessing the authenticity of such findings,
including chemical spectroscopy for analyzing and dating patinas, epigraphic
and paleographic examination by specialists in ancient semitic script, and
DNA analysis of any remains, not to mention historical scholarship of early
Christianity. Any tampering with the tombsite or other possibilities for fraud
must also be weighed and taken into account.

One purpose of this article is to contribute toward such efforts by develop-
ing statistical methods for assessing evidence for and against a "hypothesis"
that this tomb belonged to the family of the historical Jesus. In doing so we
consider such data as are available on the distribution of names during the
era in question, and we compute (on the basis of numerous assumptions de-
tailed explicitly) probabilities and estimates related to such questions as the
expected proportion of times that a similarly "surprising" sample of names
could be expected to arise by pure chance when sampling from a population
having similar characteristics to the one which existed at that time. Our
computations were carried out under a specific set of assumptions which are
by no means universally accepted. Of course, ultimately, the authenticity
of any such find cannot be determined through purely statistical reasoning
alone, and it can certainly turn out that this tombsite is not that of the
NT family; in that eventuality the validity of our *methods* should remain
unaffected. A further purpose of this paper is to lay out this highly inter-
esting data set—together with the novel inferential challenges it poses—for
the benefit of the statistical community.

In Section 2 below we describe the unearthed tomb and the ossuaries
discovered inside. Background on the practice of ossuary interment is given
in Section 3. The genealogy of the NT family—central to our analysis—is
discussed briefly in Section 4. Section 5 discusses available data sources and
provides some statistical summaries of the Jewish onomasticon, that is, of
the distribution of names of the men and women who lived during that era,
and Section 6 follows up in further detail for the particular names found
in the East Talpiot tomb. Some statistical "judgement calls" are discussed
in Section 7. Because the Talpiot tomb must be regarded as having been
"best" out of many possible observations, in Section 8 we review what is known about the size of the relevant population within which these burials took place. Section 9 addresses some inferential issues which arise in data of this type.

For statistical inference to be valid, one may not tailor an alternative hypothesis to data that has already been seen. In Section 10 we address such matters and on a best efforts basis we carefully formulate a priori hypotheses for this problem. A paradigm for the inference problem at hand is then developed in Sections 11 and 12. Our method is based on defining an a priori measure of the "surprisingness" of an observation using the "relevance and rareness" of certain name renditions, and an assumed complex of NT familial relations among them. "Relevance" will refer essentially to membership in an a priori list of *candidates* for inclusion in a NT tombsite, and "rareness" will be defined relative to an a priori list of nested possible name renditions for each such candidate; features of familial interrelations figure prominently in the formulation. Our analysis, implemented for a variety of parameter choices, is reported in Section 13 which first provides a detailed summary of the assumptions underlying our analysis. In Section 14 we provide a detailed discussion of our results, and some concluding remarks. The R computing code on which our results are based may be downloaded from the "statlib" website [Feuerverger (2008)].

We remark that, in assessing the evidence in any way, it is essential to adopt a strictly *historical* viewpoint, and thus to set aside considerations that a NT tombsite cannot exist. In fact, Jewish ritual observances prevalent at the time are entirely consistent with the possible existence of such a tomb. We caution the reader to note, however, that the analysis we present is based on one specific "tradition" of history. These assumptions represent the author's best understanding as at the time the analysis was carried out but they are far from universally agreed upon and they enter into the analysis in a cumulative way. It is anticipated that such points will be revisited in the discussion to this paper.

**2. Description of the find.**  The vestibule of the tomb was damaged by the blasting operations that led to its discovery. The tomb had otherwise been covered by earth, apparently undisturbed since antiquity. On the exterior facade above the tomb's entranceway there was carved in relief a circle beneath an upward pointing gable—a rare feature. Within the $2.3 \times 2.3$ m tomb were six *kokhim*[4]—two on each of the other three walls—each just over 1.6 m in length, and under 0.5 m in width, deep enough to store two or three ossuaries in each. Within these *kokhim* a total of ten ossuaries were found,[5] some of them broken. Two ossuary lids, discarded in antiquity, were found beneath the soil fill in the room. Early Roman (Herodian) sherds (i.e., broken pieces of pottery) were also found on the floor which date the site to the late Second Temple period, that is, from the end of the first century BCE or the beginning of the first century CE to approximately 70 CE. Such bones as were within the ossuaries were in an advanced state of disintegration. Two arcosolia (shallow shelves intended for laying out bodies) had been carved in the tomb walls and contained broken and powdered bone remains. Disturbed bones, presumably swept off the arcosolia, were also found on the floor. The *golal* (blocking stone) to the tomb's entrance was not found at the site indicating that the tomb had been accessed by robbers in antiquity.

---

[4] *Kokhim* (singular *kokh*) are small horizontal tunnels chiseled into the walls of a tomb within which ossuaries could be placed. The Latin terms are *loculus* and *loculi*.

The ossuaries found within this burial cave are typical of Jewish ossuaries of the first century CE. Six of the ten ossuaries bore inscriptions, five in Jewish script (i.e., Hebrew or Aramaic) and one in Greek. This proportion of inscribed ossuaries (i.e., 6 out of 10) and this proportion of Hebrew to Greek (5 out of 6) are both higher than typical of other tombs previously found in this area. The six inscribed ossuaries and the four uninscribed ones are described below in the order they appear in Kloner (1996); their Israel Antiquities Authority (IAA) identification numbers and dimensions are indicated as well.

OSSUARY #1. IAA 80–500. $68.5 \times 26 \times 32.5$ cm. Inscribed in Greek:

$$M\alpha\rho\iota\alpha\mu\eta\nu\sigma\upsilon \; [\eta] \; M\alpha\rho\alpha$$

This elegantly rendered ossuary (see Figure 1) has multiple possible readings. Mara, an (absolute) contracted form of (the emphatic) Martha, is a rare name, these being feminine versions derived from the Aramaic dominant masculine form mar meaning "lord," "master," or "honorable person." The question of whether Mara was intended here as a title, such as "honorable lady," or whether it was intended only as an alternate (i.e., second) name is disputed. If this inscription were understood as in Hebrew, then Mariamenou would be a diminutive (i.e., endearing) form of Mariamne or Mariamene and the inscription would read "Mariamene [diminutive] the lord/master" provided we also assume also that $M\alpha\rho\alpha$ (or מרא) is intended as "lord" or "master" and that "$\eta$" is meant as the feminine article "the." An alternate reading requires that one interpret the stroke between "Mariamenou" and "Mara" as representing not an $\eta$, but only a scratch mark; in that case one interpretation is that this ossuary contains the remains of two persons—one called Mariame, and the other called Mara. However, the manner in which these two words run closely together, and on the same line, seems more suggestive of their referring to a single person. Rahmani (1994), pages 14 and 222, reads the inscription as follows: "The stroke between the $\upsilon$ of the first and the $\mu$ of the second name probably represents an $\eta$, standing here for the usual $\eta\ \kappa\alpha\iota\ldots$ used in the case of double names…" and he posits that the second name is a contracted form [not a contraction] of "Martha" leading to the reading "Mariamene [diminutive] who is also called Mara." According to Greek usage of the time, the first word of the inscription is a genitive/possessive form for Mariamene, rendered in a particular diminutive form understood to be an endearment, so that the inscription then translates as "[the ossuary] of Mariamene [diminutive] also known as Mara." Rahmani's reading, which is the one we adopt, was accepted by Kloner (1996) and has been corroborated by others in the field.

_______________
[5]No information is available regarding the placement of the various ossuaries among the *kokhim*.

OSSUARY #2.   IAA 80–501. $55 \times 23 \times 27$ cm. Inscribed in Hebrew lettering:

<div dir="rtl">יהודה בר ישוע</div>

The lettering is executed clearly—see Figure 2. It translates as "Yehuda son of Yeshua," Yehuda being Hebrew for Judah. Note that "bar" (i.e., בר "son of") is Aramaic, not Hebrew.

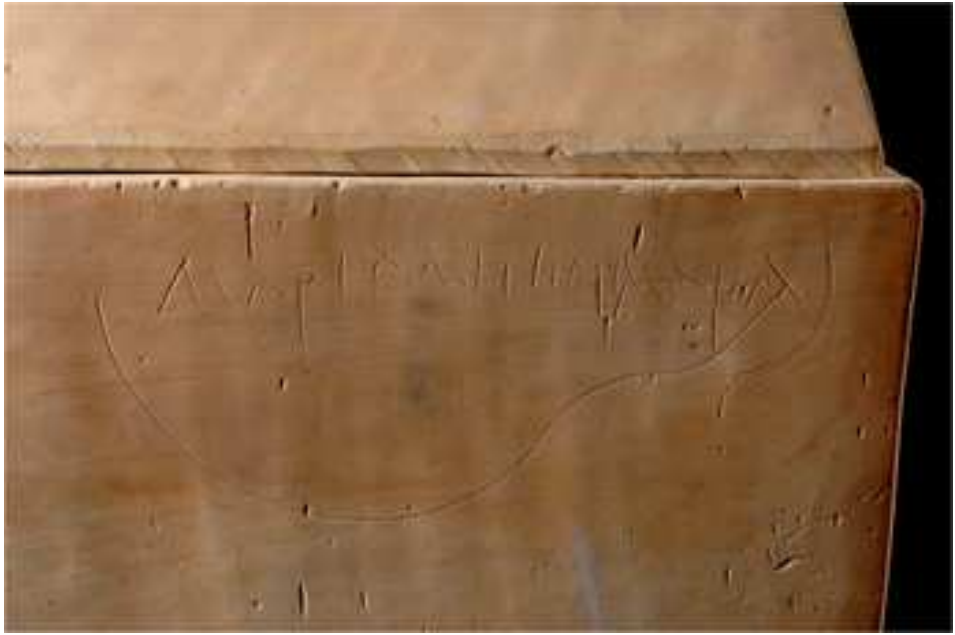

FIG. 1.

Ossuary #3.   IAA 80–502. $55 \times 28 \times 34$ cm. Inscribed in Hebrew:

מתיה

This translates as Matya, a shortened form of Mattityahu, that is, Matthew; see Figure 3.

Ossuary #4.   IAA 80–503. $65 \times 26 \times 30$ cm. Inscribed in Hebrew lettering:

ישוע בר יהוסף

This translates as Yeshua son of Yehosef, that is, Jesus son of Joseph. Unlike the other inscribed ossuaries found in this tomb, the incisions here are "messy," "informal," and superficially carved, and each of the four letters of ישוע is faint; see Figure 4. However this reading of the inscription was authenticated (by Rahmani and also Kloner) by comparison with the inscription on Ossuary #2 and is corroborated by others. Also relevant is that no other Hebrew name ends in the letters *vov* and *ayin*. A large, crudely carved rightward-leaning cross, whose purpose or symbolic meaning (if any) is unknown, appears at the head of the inscription. Cross-marks on ossuaries were sometimes carved by masons, most likely to indicate alignment of lid-tops; in this instance the marking does not have the appearance of being an *obvious* scratch mark of this nature. It has been suggested that the "cross" on this ossuary may have been purposeful.

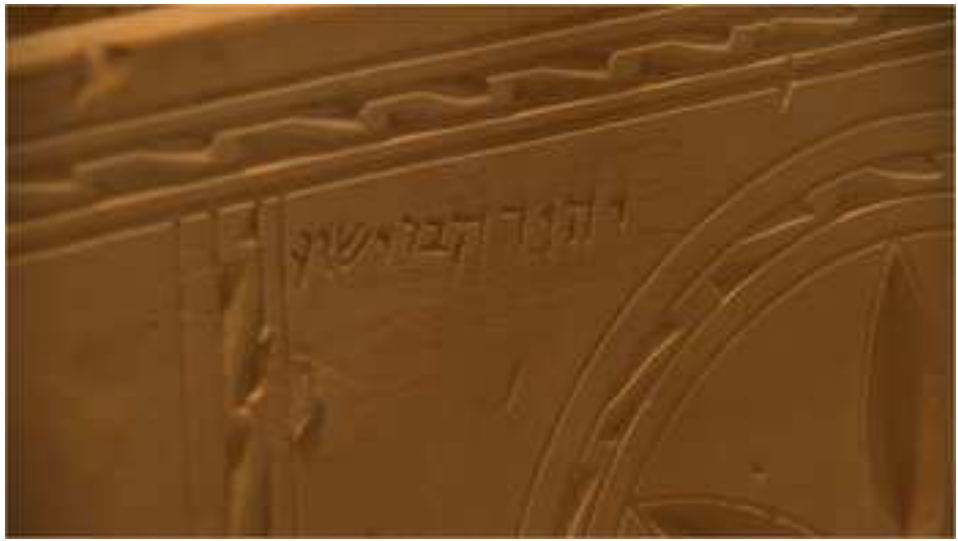

Fig. 2.

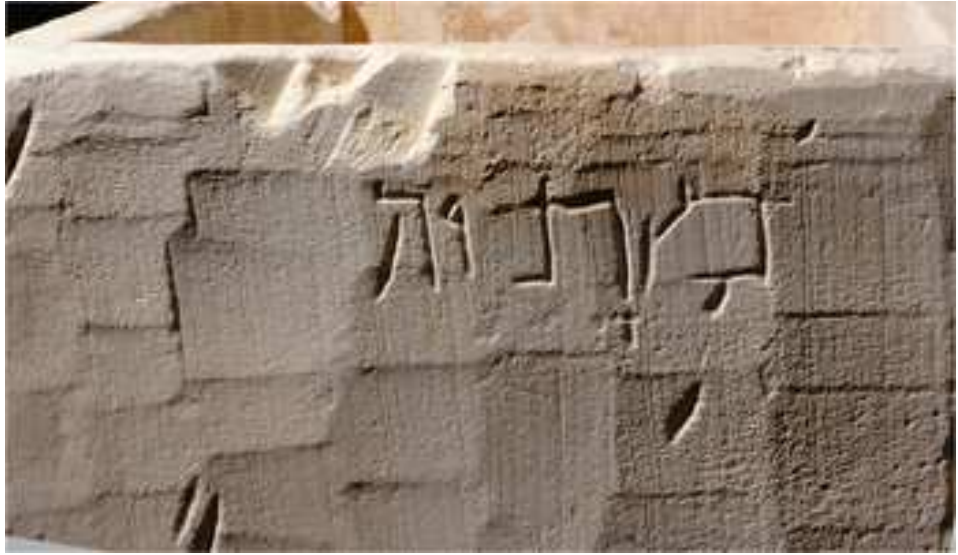

Fig. 3.

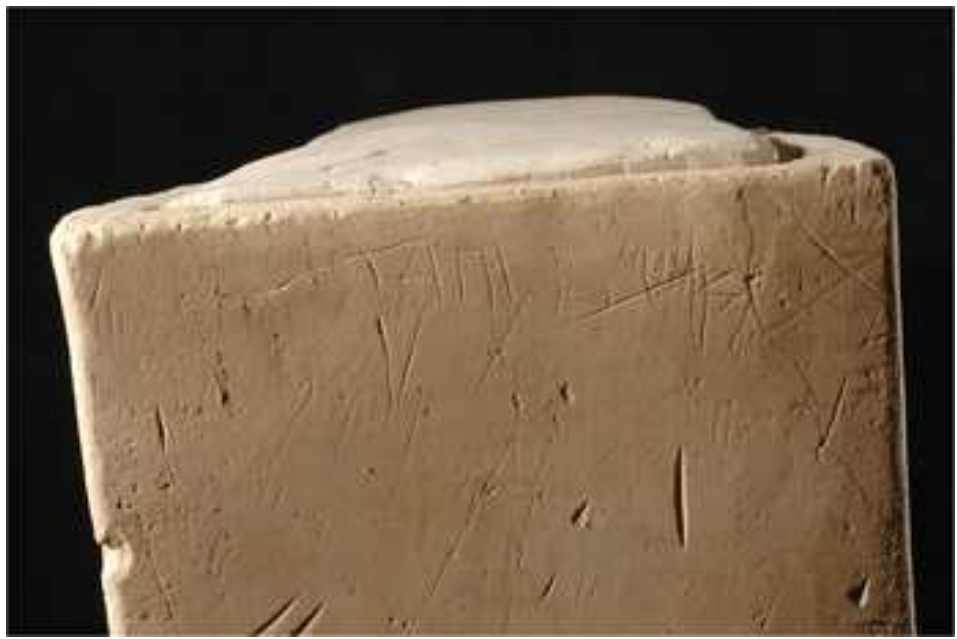

Fig. 4.

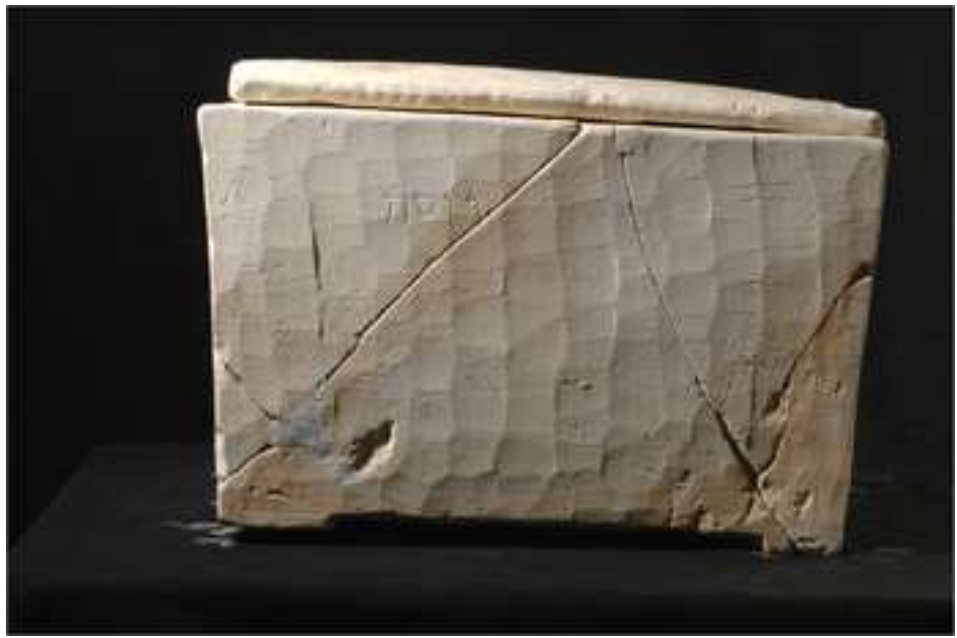

Fig. 5.

Ossuary #5.   IAA 80–504. 54.5 × 26 × 34.5 cm. Inscribed in Hebrew:

יוסה

This translates as Yoseh or Yosa, a relatively rare variant of Yosef or Yehosef (i.e., Joseph). In Hellenized form, this inscription would be read as Yose, Yoses, or Joses. See Figure 5.

Ossuary #6.   IAA 80–505. 52 × 27 × 33 cm. Inscribed in Hebrew:

מריה

This translates as Marya, that is, Maria, a Hellenized form of Miriam or Mariam. See Figure 6.

Ossuary #7–10.   These four ossuaries, the first three of which correspond to IAA numbers 80–506 to 80–508, bear no inscriptions and have dimensions 67 × 31.5 × 38.5 cm, 51 × 27 × 31.5 cm, 61 × 26.5 × 31.5 cm, and (the reported dimensions) 60 × 26 × 30 cm, respectively.

In general appearance, the six inscriptions correspond to four distinct styles. That of Yeshua is unprofessional. The ossuaries of Marya, Yoseh, and

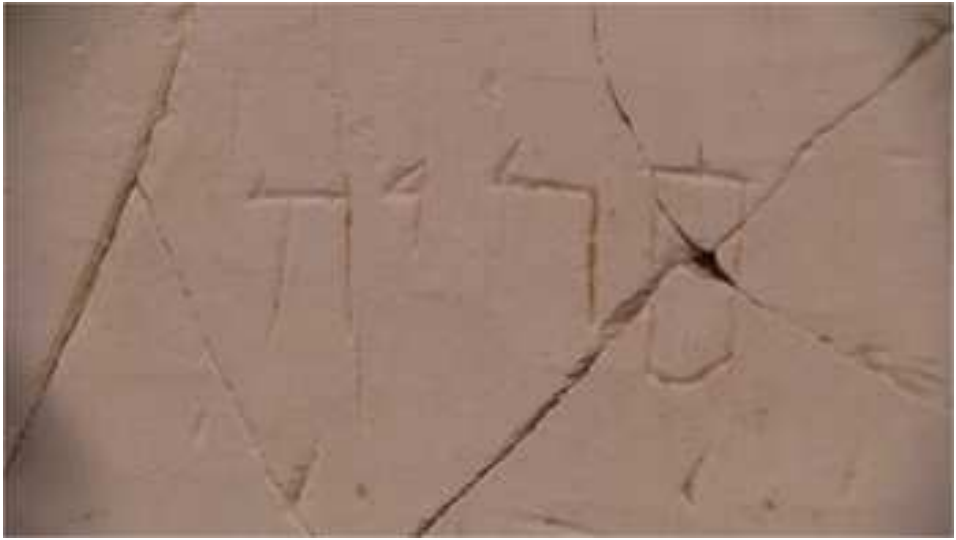

Fig. 6.

Matya are executed in similar plain but neat hands. That of Mariamenou is executed in an "elegant" Greek hand. And finally, the ossuary of Yehuda appears rendered "professionally." Rahmani surmised that the similarities between Ossuaries #5 and #6 and their inscriptions, both coming from the same tomb, may indicate that Yoseh and Marya were the parents of Yeshua and the grandparents of Yehuda.[6]

Although the dimensions of the ossuaries differ, each is consistent with the measurements of an adult. Among the inscribed ossuaries, numbers 1 and 4 (Mariamenou and Yeshua) are the longest, possibly corresponding to taller than average persons, and numbers 1 and 2 bore ornamental carvings (rosettes, etc.) as did also the first three of the four uninscribed ossuaries listed; all of the other ossuaries were ornament-free, except for such inscriptions as have been noted.

Finally, we note that the tenth ossuary—that is, the uninscribed, unornamented one with dimensions $60 \times 26 \times 30$ cm—is "missing." The original archeological drawings made at the time of the excavation indicate that ten ossuaries were found at the site, but IAA records show that only nine were retained in its permanent collections. Now, it is not entirely unusual that an ossuary—particularly an uninteresting one—would get "lost" in the comings and goings of such archeological work. However, suggestions have been raised [e.g., Tabor (2006), among others] that the dimensions of the missing ossuary seemingly match those of the disputed ossuary of James.[7] Were this so, statistical dimension matching[8] could easily be used to *prove* that the James ossuary must surely be the one missing from our tomb, with attending consequences that would be startling, particularly if the *full* inscription on the James ossuary were authenticated. Our investigations along these lines, however, did not prove fruitful.

---

[6]If this interpretation is correct, the tombsite cannot be that of the NT family. However Rahmani does not follow up with any explanation for the messy nature of the inscription on Ossuary #4.

In Sections 5 and 6 we shall discuss the distribution of Jewish names in late antiquity and provide some further details concerning the names found on the Ossuaries #1 through #6. The next two sections provide some background on the practice of ossuary burial, and on the genealogy of the NT family.

**3. Ossuaries and re-interment.** An ossuary is an approximately rectangular chest, typically quarried in the soft limestones common near Jerusalem, containing the bones of one (and sometimes more) deceased persons. The custom of repositing bones of the dead in such bone boxes is not mandated by *halacha*, that is, Jewish ritual law; it was practiced by Jews in and around Jerusalem only from the end of the first century BCE, or from the start of the first century CE, until the year 70 CE. Instead of burial in coffins as had been an earlier custom, bodies were apparently first placed in a pit or a cave and left to decompose for about a year until only bones remained. These bones were then gathered by the deceased's family, deposited into an ossuary, and interred in a tomb. Ossuaries (and tombs in particular) were a more costly form of burial that not all persons could afford. Further information and speculation regarding the religious and politico-historical aspects of this practice, may be found in Hachlili (1994), Kloner (1996) and Rahmani (1994).

The approximate dimensions of ossuaries are usually recorded in centimeters in the order length × width × height. Typical ossuary boxes are somewhat tapered so that the length × width dimensions at the top will be slightly larger than at the bottom. Being quarried and chiseled artifacts, the shape, and hence the dimensions, of ossuary boxes are not entirely precise. The length of an ossuary had to be sufficient to house the femur (thighbone) which is the human body's longest bone, and the two other dimensions had to be sufficient to house the skull, pelvis, and other bones.

---

[7]An ossuary inscribed "James son of Joseph brother of Jesus" is in the possession of Oded Golan, a private Israeli antiquities collector, under prosecution for alleged forgery at the time this article was written; see, for example, Shanks and Witherington (2003). Israeli prosecutors apparently accept the authenticity of the first component of this inscription but allege that the second component had been forged, although (as of the time of writing) no evidence to that effect has been produced. The statistical aspects of the "James son of Joseph brother of Jesus" inscription were studied by Fuchs (2004).

[8]Rahmani (1994) gives the dimensions of a sample of 897 ossuaries from which the multivariate distribution of dimensions can, for this purpose, be quite reliably inferred.

Ossuaries were frequently carved with ornamental motifs such as lattices, friezes, triglyphs, or rosettes. Such markings would typically have helped identify the persons lying within, especially for ossuaries that were uninscribed (as might occur, e.g., in families lacking literacy).

Rahmani (1994) notes that 233 of the 897 ossuaries in the State of Israel collections as of 1989 bear inscriptions[9] meant to identify the individuals within, with these inscriptions being in one or more of the languages in common use at the time—primarily Aramaic, Hebrew and Greek. About two thirds of these inscriptions are in Hebrew/Aramaic, while about one third are in Greek, or a combination of Greek and Hebrew. The use of inscriptions evidence some degree of literacy on the part of the family to whom the tomb complex belonged. In virtually all cases, such inscriptions consist only of the individual's first name,[10] or of their first name together with the name of their father. Inscriptions for women occasionally included the name of the husband in lieu of the father. Only a single case among the ossuaries catalogued by Rahmani includes the name of a brother, and only one mentions the name of a son; such rare mentions presumably occurred only when the other mentioned persons were individuals of particular distinction. Contractions of names appear also to have been used, and were likely intended as endearments. Note that the use of inscriptions was intended solely to assist members of the immediate family to identify the remains within; they served no public or other function.

Ossuary burial was practiced primarily within the environs of Jerusalem in part, no doubt, because of the availability of suitable stone there. In fact [Ilan (2002), page 52], of the 712 names in Ilan derived from ossuaries sources, only 66 were found outside of the Jerusalem region, with 24 of these having come from a single burial cave in Jericho. Rahmani (1994), page 21, notes that ossuaries quarried at Jerusalem were also used by Jews living as far away as 25 km from Jerusalem (including Jericho).

## 4. A brief NT genealogy.

The names in the genealogy of the NT family bear on the statistical analysis; however, our discussion here will be brief. We caution the reader that our analysis relies on a specific "tradition" for this geneology but that such historical details cannot be regarded as being certain.

---

[9]Because plain ossuaries are of lesser interest and often become "discarded," these figures significantly overstate the inscription rate.

[10]According to *halachah* the name marked on a grave must correspond to the actual name by which that person was known during their lifetime. In particular, if an individual had been known by a nickname, that form of their name must be used on their coffin. Note that although *halachah* postdates the era of Jesus we are assuming here that this basic tenet was already essentially being observed at that time.

Jesus was born a few years before the turn of the millennium and was crucified in (most likely) April of 30 CE. The earliest known historical record of the names of Jesus' siblings is provided by Mark 6:3 (written around 70 CE) who lists the names of Jesus' brothers in the order James, Joses, Judah, and Simon. Since it was customary to name the eldest first, it is reasonable to assume that James was Jesus' eldest brother, and Joses was next eldest. Matthew 13:54–56, usually believed to be a historically later source, records the names in the order James, Joseph, Simon, and Judah—using Joseph in place of Joses for the second brother, and reversing the order of the last two names. It seems likely that Judah was actually the youngest, for upon Jesus' death James took over the ministry, and upon James' death Simon (and not Judah) did—Joses thus having likely no longer been alive at the time. These sources also refer to sisters of Jesus in the plural but do not name them.

The earliest extant versions of Mark and Matthew were originally written in Greek, with Mark being considered here to be the earlier and therefore more authoritative source. Hence the earliest known written record refers to the second brother as Joses, and not as Joseph. We shall take Joses as having been the actual name of that brother.

It is commonly believed that Jesus had two sisters and that they were called Mariam and Salome. A single (and later) source whose reliability seems less certain suggests there may have been a third sister named Joanna.

Joseph was the son of Jacob (i.e., Yaakov, or James), and 2nd century sources name the parents of Mary as Joachim and Anna of Sepphoris[11]—the largest city in the vicinity of Nazareth at the time. Concerning further ancestry, at the start of the NT there is a lengthy series of "begats" (i.e., geneological lists) whose purpose is to trace the lineages of Mary and Joseph back to King David; these can arguably be used to study their genealogies. In particular, the name Matya appears several times in the lineage of Jesus (as recorded in Luke) and some scholars attribute this name to the lineage of Mary. The tracing of ancestries back to the house of David relates to the theme of the NT since it may have been commonly held that the lineage of the Messiah would trace back along a "Davidic line."

Concerning the ultimate fate of the siblings of Jesus, only a small amount is known. Paul 1, Cor. 9:1 ff refers to the brothers as traveling with their wives which suggests that they were married and likely had children. The names of these women and any children are not known, although a reference is known to grandsons of Yehuda named Zoker and James.

---

[11]Sepphoris was savagely destroyed by the Romans in 4 BCE and later rebuilt by Herod Antipas.

Josephus Flavius (1943) records the execution of James in 62 CE in the vicinity of the Temple, stating that this James was "the brother of the man known as Jesus who is called the Messiah." Consequently, James may be regarded as an a priori candidate whom one might not be surprised to find in a NT family tomb, if one such existed (although early historical records appear to indicate that James was buried at the place of execution). The two youngest brothers Simon and Judah are both surmised to have lived beyond the year 70 CE, into the reign of Trajan (tenth emperor of Rome who ruled between 98–117 CE) and are therefore not a priori candidates for such a tomb. The fate of Joses is unknown; after he is mentioned by name in the gospels he is never heard of again. However, because it was Simon who succeeded as leader when James died, it is generally assumed that Joses was no longer alive at the time. Joses is therefore an a priori candidate for a NT tomb. As for Judah, the manner of his death is not known.

Concerning any possible "wife" of Jesus, nothing is known except that had one existed she would likely have been interred in the family tomb if there were one.[12] Jesus too is, of course, a candidate for a NT tombsite, and we also know—from the NT passages concerning Joseph of Arimathea—that persons who pre-deceased Jesus are not candidates for such a tombsite since the family evidently did not possess one prior to Jesus' death.

**5. Statistics of the Jewish onomasticon.**  At least three resources are available for studying the distribution of names during the era relevant to this study. The first is the catalogue of Jewish ossuaries in the collections of the State of Israel compiled by Rahmani (1994) who details all ornamented and inscribed ossuaries held by the Israel Antiquities Authorities (IAA) and by the Israel Museum as of 1989—a total of 897 specimens in all. Of these, 233 bear inscriptions identifying the names of a total of 241 male persons and a substantially lesser (but undetermined) number of female persons. Of the 233 inscribed ossuaries, 143 are in Jewish script (i.e., Hebrew or Aramaic), 73 in Greek script, and the remainder in a mix of both scripts or in other languages (such as Latin). A total of 147 unique *names* (male and female) occur among them. The compilation in Rahmani is not arranged by either tomb groups or by gender, and only limited summary information is provided on the distribution of names. Although it is, in principle, possible to do so by working with an index of names provided, it is not straightforward to abstract statistical information from this source.

The second resource, and by far the most comprehensive one currently available, is the lexicon of Jewish names of late antiquity compiled by Tal

---

[12]The only "viable" *candidate* for a "wife," assuming one existed, appears to be Mary Magdalene although we shall make no such assumption. Mary Magdalene does, however, turn out to be an a priori candidate for inclusion in a NT tombsite based on other grounds.

Ilan (2002). It covers the period between 330 BCE (marking the Hellenistic conquest of Palestine) and 200 CE (which marks the closing of the Mishnaic period and of the early Roman Empire). Ilan's compilation includes the names of 2509 males and 317 females taken from all available sources, including not only ossuaries from both within as well as outside the State of Israel collections, but also from literary sources, epigraphic and papyrological documents, and many other sources. Detailed source information and some statistical compilations are also provided. Although Ilan includes all recorded names used by Jews of Palestine during the stated period, it also includes a further 86 names of women and 685 names of men regarded as fictitious, that is, not corresponding to persons who had actually lived. Fictitious names will be excluded from our analysis.

A third resource is Hachlili (2005); in particular, Tables V-2, (a) and (b) of Hachlili (page 200) provide frequencies for the most common personal names among Jews, by gender and by source, in the late Second Temple period. These tabulations are based only on the most common names—for a total of 1091 males and 192 females—taken from ossuaries, Masada ostraca, and other sources. The sample sizes of which these common names constitute subsets are not provided. These tables essentially coincide with subsets of names in Ilan (2002) but dating to the late Second Temple period.

Ilan's more extensive compilation allows less variable estimation of the incidence of names, although estimates meant to pertain only to the population of ossuaries, but based on all of Ilan, may be somewhat biased not only because nonossuary sources are thereby included, but also because Ilan's compilation includes periods some 300 years prior to when ossuary burials became prevalent as well as 130 years after that practice had ceased. Estimates based on the samples of Rahmani or Hachlili will be much more variable, but presumably less biased, based as they are, in the first instance, on names appearing on actual ossuaries only, and in the second, on names from the late Second Temple period only. It is possible to extract from Ilan's lexicon names obtained only from ossuaries, and these constitute a superset of the sample in Rahmani. Of course, one could argue that no population assembled from such sources can be regarded as valid for the inference at hand, however, we regard that viewpoint as nihilistic and shall not adopt it.

Although the information in Ilan (2002) is not arranged specifically for our purposes, the compilations there include names taken from ossuaries as well as from many other sources, and further, many more names taken from ossuaries appear in Ilan than in Rahmani since Rahmani catalogues only ossuaries in the State of Israel collections while Ilan includes names on ossuaries from all available sources. As already mentioned, Ilan contains the names of 2509 male persons and 317 female persons. These comprise 721 unique male names and 110 unique female names. Furthermore, Ilan states that, of these, the names of 519 male and 193 female persons (712 persons in all) had been derived from ossuary inscriptions (numbers substantially higher than Rahmani). From this it *appears* that about 27% of inscribed ossuaries bear female names, while 73% bear the names only of males; however the relative frequency of ossuaries of females is underrepresented in these numbers due to the custom of sometimes naming fathers on both male as well as female ossuaries, and of occasionally naming husbands on female ossuaries. Note also that 61% of the female names in Ilan are derived from ossuary sources while only 21% of the male names are so derived, numbers that reflect the patriarchal nature of society at the time.

Our presentation of these distributions of names is laid out in Tables 1 through 5. Table 1 gives the total number of unique male and female *persons* in each of Ilan and Rahmani, as well as the corresponding number of unique male and female *names*. The fourth column gives Ilan's counts when restricted to names obtained only from ossuary inscriptions. In this table, as in some of the others below, not all tabulations or computations were completed, either for reasons of feasibility or for constraints of time; this will be indicated throughout by dashed lines at the affected table entry positions. It will be important to bear in mind that *dashes in the tables do not represent zeros.*

Table 2 gives the ten most common female names according to Ilan, together with their frequencies in Ilan, Rahmani, and among Ilan's ossuary sources only. There are (very) slight variations between the numbers in our table and a similar one in Ilan, ours having been corrected for a small number of additional entries Ilan had later added to her lexicon. Fictitious name counts are shown separately, with "F" labels attached; for example, Ilan lists 63 Salomes, but two were fictitious. Note that names obtained from ossuaries are never fictitious. Here again dashed lines represent undetermined entries (not zeros).

Table 3 gives the 21 most common male names appearing in Ilan, together with their frequencies in Ilan, Rahmani, and among Ilan's ossuary sources only, with slight updates having again been made to a similar table of Ilan. There are also minor differences between the Rahmani column of our table, as determined by us, and a table based on Rahmani given by Fuchs (2004). The fictitious name counts in the "Ilan" column again occur only on nonossuary sources; in one instance (an Eleazar) the fictional status is uncertain.

TABLE 1
*Onomastic gender distribution*

| Gender | Ilan | Rahmani | Ilan ossuaries |
|---|---|---|---|
| Male persons | 2509 | 241 | 519 |
| Female persons | 317 | – | 193 |
| Total persons | 2826 | – | 712 |
| Male names | 721 | – | – |
| Female names | 110 | – | – |
| Total names | 831 | 147 | – |

TABLE 2
*Jewish female names of late antiquity*

| Generic name | Ilan | Rahmani | Ilan ossuaries |
|---|---|---|---|
| Mariam/Mary | 74 + 6F | 18 | 44 |
| Salome | 61 + 2F | – | 41 |
| Shelamzion | 25 + 0F | – | 19 |
| Martha | 21 + 0F | – | 17 |
| Joanna | 12 + 0F | – | 7 |
| Shiphra | 12 + 0F | – | 9 |
| Berenice | 9 + 1F | – | 2 |
| Sarah | 8 + 1F | – | 5 |
| Imma | 8 + 0F | – | 6 |
| Mara | 7 + 0F | 2 | 5 |
| No. females | 317 + 86F | – | 193 |
| No. female names | 110 | – | – |

A number of difficulties occur in producing such tables. In Rahmani (1994), the gender of several of the names is ambiguous. (Presumably one could try to resolve these by cross-referencing to Ilan where most names are categorized by gender.) Furthermore, some inscriptions are uncertain due to problems of legibility. The resulting tables therefore depend somewhat on what conventions one adopts toward the various problems of this nature.

Ilan (2002) and Hachlili (2005) give considerable further information concerning the customs of naming as well as about the distribution of names in that era. By way of general comment, one can say that the pool of names in use was not unlimited. For that reason different renditions of a generic name category often acted as distinct names so as to help distinguish among individuals. Names associated with the Hasmonean dynasty were especially popular. For men, these include the names Mattathias, Yochanan, Simon, Judah, Eleazar, and Yonatan. As for Hasmonean women, only two of their Hebrew names are known—one called Mariam, and the other Shelamzion. It is possible that the name Salome was popular for the same reason, but its origin is uncertain. Biblical names, particularly of the secondary characters, were also popular, with the names of primary biblical characters being less prevalent than might have been expected.

TABLE 3
*Jewish male names of late antiquity*

| Generic name | Ilan | Rahmani | Ilan ossuaries |
|---|---|---|---|
| Shimon/Simon/Peter | 249 + 8F | 24 | 62 |
| Yehosef/Yosef/Joseph | 221 + 10F | 19 | 45 |
| Yehudah/Judah/Judas | 171 + 8F | 20 | 45 |
| Eleazar/Lazarus | 169 + 7F + 1? | 14 | 30 |
| Yochanan/John | 124 + 5F | 8 | 26 |
| Yehoshua/Yeshua/Jesus | 101 + 2F | 10 or 11 | 23 |
| Hananiah/Ananias | 83 + 3F | 11 | 19 |
| Yonathan/John | 72 + 3F | 6 | 14 |
| Mattathias/Matthew | 62 + 1F | 7 | 17 |
| Menachem | 44 + 2F | 0 | 4 |
| Yaakov/Jacob/James | 43 + 2F | 5 | 6 |
| Hanan | 36 + 3F | 4 | 7 |
| Alexander | 30 + 1F | 4 | – |
| Dositheus | 30 + 1F | 6 | – |
| Zachariah | 25 + 6F | 1 | – |
| Ishmael | 31 + 0F | 2 | – |
| Levi | 25 + 4F | 1 | – |
| Saul | 29 + 0F | 10 | – |
| Choni/Onias | 27 + 0F | 0 | – |
| Shmuel/Samuel | 21 + 5F | 0 | – |
| Hezekiah | 23 + 3F | 0 | – |
| No. of males | 2509 + 685F | 241 | 519 |
| No. of male names | 721 | – | – |

The counts shown for each of the *generic* names in Tables 2 and 3 include all *renditions* or variants of that name. However, we shall require more detailed statistical information regarding the *renditions* within the *generic* categories for certain names relevant to this study. Three variants will interest us particularly, namely the variants Mariamenou and Marya for Mariam, and the variant Yoseh for Yosef. Such breakdowns are provided in Tables 4 and 5. We see from Table 4 that there are (in all) 16 variants for Mariam, and from Table 5 that there are 22 variants for Joseph if language differences are also allowed for. In Table 4, horizontal lines demark two groups of Mariam renditions relevant for us, with Mariamenou and Mariamne isolated at the top of the table and versions "equivalent" to Marya isolated at the bottom; close-sounding versions are placed close to, but on the opposite sides, of these lines. Likewise, in Table 5, the renditions considered relevant to the biblical brother Joses appear in the five rows isolated at the bottom.

We note the following important differences between ossuary and nonossuary sources. For Mariam, the rendition מריה apparently occurs only on ossuaries. For renditions of Joseph, the form יוסי never appears on ossuaries, while the Greek form $I\omega\sigma\eta\pi\sigma\varsigma$ and the Hebrew form יוסף are also greatly underrepresented on ossuaries. The rendition יהוסף is the most common one appearing on ossuaries, although it is well represented among nonossuary sources as well. In the five renditions (at the bottom of Table 5) consistent with the biblical brother, their "free use" on ossuaries, and relative rareness on nonossuaries, appears consistent with the notion that they act much like a separate name category. Of these five, the Hebrew rendition יוסה has never been found on any ossuary other than at Talpiyot.

TABLE 4
*Mariam renditions*

| Rendition of name | Ossuary sources | Nonossuary sources | Combined sources |
|---|---|---|---|
| $M\alpha\rho\iota\alpha\mu\eta\nu\sigma\upsilon$ | 1 | 0 | 1 |
| $M\alpha\rho\iota\alpha\mu\eta$ | 0 | $0 + 1F$ | $0 + 1F$ |
| $M\alpha\rho\iota\alpha\mu\eta\nu$ | 0 | 1 | 1 |
| $M\alpha\rho\iota\alpha\mu$ | 2 | 2 | 4 |
| $M\alpha\rho\iota\alpha\mu\eta$ | 10 | 8 | 18 |
| $M\alpha\rho\alpha\mu\eta$ | 0 | 1 | 1 |
| $M\alpha\rho\iota\alpha\mu\eta\varsigma$ | 0 | 1 | 1 |
| $M\alpha\rho\iota\alpha\delta\sigma\varsigma$ | 1 | 0 | 1 |
| $M\alpha\rho\iota\epsilon\alpha\mu\eta$ | 1 | 0 | 1 |
| מרים | 12 | $10 + 4F$ | $22 + 4F$ |
| מרים | 3 | 0 | 3 |
| מרימא | 0 | 1 | 1 |
| $M\alpha[\rho]\iota\alpha\varsigma$ | 1 | 0 | 1 |
| $M\alpha\rho\iota\alpha$ | 4 | $6 + 1F$ | $10 + 1F$ |
| מריה | 8 | 0 | 8 |
| סריה | 1 | 0 | 1 |
| Total of above | 44 | $30 + 6F$ | $74 + 6F$ |

**6. More about the Talpiyot inscriptions.** In this section we provide some further details for the particular names occurring on the Ossuaries #1–6 described in Section 2. Our primary resource here, again, is Ilan (2002).

**Mariam & Marya:** The name Mariam or Miriam, and its variants, was the most common female name of the Second Temple era.[13] We note also that starting with the earliest gospels of Mark, Marya is the principal form by which the name of the historical Mary has been handed down; it is therefore likely that this is the form of the name by which she was known. (We remark that this contention is not universally accepted.)

---

[13] We are following the statistics of Ilan's onomasticon here; some sources put Salome as the most common female name, with Mariam as the second most common.

Table 5
*Joseph renditions*

| Rendition of name | Ossuary sources | Nonossuary sources | Combined sources |
|---|---|---|---|
| $I\omega\sigma\eta\varphi o\varsigma$ | 0 | $0 + 5F$ | $0 + 5F$ |
| $I\omega\sigma\eta\pi o\varsigma$ | 4 | 38 | 42 |
| $I\omega\sigma\iota\pi o\varsigma$ | 0 | 1 | 1 |
| $I\omega\sigma\eta\pi o\upsilon$ | 1 | 6 | 7 |
| $I\omega\sigma\eta\varphi$ | 2 | 6 | 8 |
| $I\omega\sigma\eta\pi$ | 0 | 3 | 3 |
| $I\omega\sigma\iota\alpha\varsigma$ | 0 | 1 | 1 |
| $I\omega\sigma\iota o\upsilon$ | 0 | 1 | 1 |
| Ioseph | 0 | $0 + 1F$ | $0 + 1F$ |
| Iosepu | 0 | 1 | 1 |
| יוסף | 2 | $17 + 2F$ | $19 + 2F$ |
| יהוסף | 27 | 61 | 88 |
| יהוסה | 1 | 0 | 1 |
| יהסף | 2 | 1 | 3 |
| יוסי | 0 | $29 + 2F$ | $29 + 2F$ |
| איסי | 0 | 6 | 6 |
| אסי | 0 | 1 | 1 |
| $I\omega\sigma\eta$ | 1 | 1 | 2 |
| $I\omega\sigma\epsilon$ | 2 | 0 | 2 |
| $I\omega\sigma\eta\varsigma$ | 2 | 0 | 2 |
| יסה | 1 | 0 | 1 |
| יוסה | 1 | 2 | 3 |
| Total of above | 46 | $175 + 10F$ | $221 + 10F$ |

**Mariamenou [$\eta$] Mara:** Of the occurrences of the generic Mariam in Ilan (2002) only one instance consists of the "full" and highly unusual form $M\alpha\rho\iota\alpha\mu\eta\nu o\upsilon$; it corresponds to our Ossuary #1 on which the additional detail "[$\eta$] $M\alpha\rho\alpha$" is inscribed. The form $M\alpha\rho\iota\alpha\mu\eta$ also occurred only once but does not correspond to a person who actually lived, while $M\alpha\rho\iota\alpha\mu\eta\nu$ also occurred once, although not on an ossuary. We remark that Mariamenou and Mara are each individually quite rare names so that either of these should have sufficed for purposes of identification by family members if referring to a single individual.

An argument can be put forth that the actual name of Mary Magdalene was Mariamne. For some background, we refer to Bovon (2002) and references therein. In a 4th century version of the Acts of Philip, a woman who is thought to be Mary Magdalene is referred to throughout as Mariamne, and Bovon surmises that Philip was her brother.[14] This version of these Acts is the earliest and most complete one known and is also one of the earliest known historical sources explicitly citing Mary Magdalene's name. These Acts also indicate that she died in Palestine, thus potentially allowing that an ossuary of hers might be found in Jerusalem. James Tabor [private communication] has recently found a still earlier reference. Hippolytus, a second century Christian writer, wrote in *Refutations* 5.2: "These are the heads of very numerous discourses which the Nassenes assert that James the brother of the Lord handed down to Mariamne." This reference dates to approximately 175 CE, some 100 years after the destruction of Jerusalem, and furthermore suggests that "Mariamne" was, at one time, the head of a ministry thereby entitling her to be addressed as "lord" or "honorable lady." The family buried at Talpiyot appears to have understood Aramaic over a period of some two generations (in view of their use of בר) and is therefore likely to have known the Aramaic meaning of "mara."

As her name indicates, Mary Magdalene came from Magdala (or Migdal); she herself likely spoke Greek and is believed by some to have preached extensively among Greek-speaking Jews. It has been speculated that she was also an apostle and a key contributor to the early Christian movement, and explanations have been advanced (revolving around patriarchal intrigues) as to why she may have later been portrayed as a "sinner." Ossuary #1 is the only one in the Talpiyot tomb in Greek script. Since Mary Magdalene was not a descendent of the same bloodlines as the family of Jesus, it is at least plausible—if this really were her ossuary—that it might have been rendered in Greek script even while the others may not have been. The inscription on Ossuary #1 will be regarded in our analysis as an appropriate rendering of her name. As an inscription, Mariamenou [η] Mara is extraordinary, and—all previous considerations aside—among the 74 Mariams whose names are currently known to us, it provides arguably the "closest fit" to Mary Magdalene.

Our analysis will be based on the following specific assumptions concerning the inscription on Ossuary #1: First, we will assume that it refers to only one person and that it represents an appropriate appellation for Mary Magdalene. Second, we will assume that this rare rendition is not applicable to many other "Mariams." Further—inferring from the remarkable detail of this inscription—we will assume that even if a larger sample of Mariams could somehow be obtained, it is unlikely that so *specifically* appropriate a name (for Mary Magdalene) would arise with frequency greater than occurs in Ilan's sample. The reader should note that these assumptions are far from universally accepted. We shall revisit this matter in Section 14.

---

[14] The mentioned "argument" then only requires us to assume that a brother would know his own sister's name.

**Yeshua:** The name Yeshua is a derivative of Yehoshua and is the sixth most common Jewish male name of the Hellenistic and Roman periods. Its popularity derives from the fact that Yehoshua was the successor to Moses. Note that the shortened form Yeshua is the one by which the name of Jesus is known, and all literary records—whether based on the NT or on its Hebrew versions—use that form for the name. Jesus quite likely preached in Aramaic and is, in any case, known to have been able to speak it; in this respect, the use of Aramaic on Ossuary #4 is therefore not implausible.

**Yehosef & Yoseh:** The name Yehosef was the second most common male name in the Second temple period. The form Yoseh which appears on Ossuary #5, however, is an uncommon version for this name. Among the 46 ossuaries bearing some version of the name Yehosef, only one (corresponding to our Ossuary #5) bears the Hebrew form יוסה; furthermore, this version of the name is one that corresponds to that used in the gospel of Mark.[15] In our analysis, we will assume that the (father) Yehosef named on Ossuary #4 is not the same individual as the Yoseh named on Ossuary #5, and that the two name versions were intended for deliberate distinction. The rationale behind this lies, first, in the seemingly special characteristics of the name יוסה, and second, in the fact that *halacha* (although a later tradition) mandates that the name by which a person was actually known in life is the form that must appear on their gravesite. Third, the use of the somewhat informal Yeshua (instead of the more formal Yehoshua) in the patronym of the Yehuda ossuary suggests that the Talpiyot tomb family may have respected "nicknames." We note again, however, that these assumptions are not universally accepted.

**Matya:** This is a shortened form of Matityahu (Matthew), a common name having Maccabean and Hasmonean origins. According to Luke and Matthew, this name occurs in the genealogy of Jesus several times, through Mary's lineage in particular.

**Yehuda:** This translates as Judah, a strong Maccabean name, and the third most common Jewish name in the Hellenistic and Roman periods. It is also the name of a younger brother (or half-brother) of Jesus.

**7. Some statistical "judgement calls."**  In this section we indicate some statistical "judgement calls" and approximations which we propose to apply. The first is a specialized assumption concerning the independence of assignment of names. In particular, we shall assume that fathers called Yehosef

---

[15] In the earliest extant (Greek) version of Mark, the name of the brother Joses is written only as $I\omega\sigma\epsilon$ or as $I\omega\sigma\eta\sigma$. It translates into Hebrew pronounced as Yoseh (rather than Yosa).

would name a son Yeshua with frequency comparable to that in the general
population (although subject to the proviso that the names of fathers and
sons ought normally to differ); likewise, we shall assume that men called
Yehosef would marry women called Mariam in the same frequency as that
name occurs generally; and so on. Assignment of names within families is
well known to be dependent *time-longitudinally,* with children frequently
named after earlier "nodes" on their family tree. However in the present
context this assumption is applied primarily on a *time-cross-sectional* basis.
Although this assumption is unlikely to be accurate with respect to very rare
and/or very unusual names, for the types of names which concern us the di-
mension of the underlying distribution here seems small enough that modest
time-cross-sectional dependencies should not have excessive impact. Much
as we would prefer to avoid such an assumption, an incisive analysis without
it does not seem feasible. We shall, however, revisit this in Section 14.

We shall also occasionally ignore certain small (and generally negligible)
corrections to joint frequencies for such facts as that brothers ought nor-
mally to bear different names, and so on. In contexts where these could
matter more substantively (as in our computing code [Feuerverger (2008)],
for example) appropriate corrections will be taken into account.

We next address the question of biases in the samples available for as-
sessing the name frequencies. We first consider the situation for the generic
name categories and afterward for the renditions occurring within them.
There are several potential sources of bias if Ilan's complete lexicon is used.
One is the usual selection bias relating to representativeness of the sources.
Difficulties of this type affect many surveys and here little can be done to
correct them.

Another source of bias arises if nonossuary listings are included in the
frequencies. One may attempt to address this (for the generic names) by
comparing their frequencies by ossuary and nonossuary sources; these may
be determined from the second and fourth columns of Tables 2 and 3. Such
comparisons do not suggest biases of great consequence; tests for the equality
of proportions between ossuary and nonossuary sources proved to be non-
significant, although among generic names not relevant to this study there
are one or two instances among the more unusual names where the relative
frequencies between ossuary and nonossuary sources appear to differ more
substantively. As it seems preferable to allow some element of bias in return
for reduced variability (in hope of obtaining estimates with smaller overall
error) we shall use Ilan's full lexicon to estimate the relative frequencies for
the *generic* name categories relevant to our work.

As will be evident later, *smaller* frequencies for *relevant* names *in-sample*
are "advantageous" for driving tests toward "significance," while *smaller*
frequencies for *relevant* names *out-of-sample* will drive tests away from
significance. In these respects, the frequencies for such names as Simon,

Yehudah, and Matthew will ultimately not matter for us, and those for the names Joanna and Martha will matter rather little. For Mariam, Salome, and Joseph, the combined versus the ossuaries-only relative frequencies are essentially identical. For Yeshua and Yaakov the frequency differences each fall in their nonconservative directions although not significantly so, and the effects of this can be studied in experimentation.

A third source of bias stems from the fact that Ilan's lexicon covers a broader range of dates than relevant for us, this being the case (although very much less so) even if Ilan's data were restricted to ossuary sources alone. One could, in principle, study this effect by laboriously categorizing the individual entries in Ilan, however the ossuary versus nonossuary comparisons do already largely address this concern.

For the *renditions* of names within the *generic* categories the situation is, however, altogether different as Tables 4 and 5 have shown, presumably reflecting variations in the popularity of specific renditions over time. Allowances for this are necessary. To obtain estimates for name *rendition* frequencies we propose to use the overall proportion (i.e., including nonossuaries) for the generic categories—these being judged the most stable in terms of bias-variance tradeoff—but to correct "internally" for differences in the ossuary versus nonossuary rendition frequencies. Thus for the rendition Yoseh, we estimate its frequency as

$$\frac{(7/46) \times 221}{2509} = \frac{33.63}{2509},$$

since there are 221 (nonfictitious) Josephs among Ilan's 2509 males, while among the 46 Josephs whose names are derived from ossuaries, 7 were versions deemed consistent with Yoseh. Note that the frequency derived above is considerably higher (hence more conservative) than the value 10/2509 obtained "directly." Likewise the frequency for Marya will be estimated as

$$\frac{(13/44) \times 74}{317} = \frac{21.86}{317},$$

and not as 19/317, and so on. Needless to say, it is the fraction from within the generic categories that will primarily drive the variability of such estimates.

**8. Size of the relevant population.** We require estimates of the size of the relevant population of Jerusalem and of the number of ossuary burials that took place overall. The estimates in this section draw on various sources. In particular, in a paper on the James ossuary, Camil Fuchs (2004) carefully estimated the population of Jerusalem in a sequence of steps which we summarize here.

First, citing studies by Hachlili (1994) and Kloner (1980), Fuchs notes that the maximum range of dates during which Jews practiced ossuary burial was between 20 BCE and 70 CE, an interval of approximately 90 years. These, however, are outside limits, and since the practice of ossuary burial was undoubtedly introduced gradually, a reasonable, but still conservative estimate, is to assume that the custom was prevalent between 6 CE and 70 CE, an interval of some 65 years.

Second, citing studies by Broshi (1977, 1978) and Levine (2002) who estimate the habitable areas of Jerusalem and their population densities, and the study by Wilkinson (1974) on the capacity of water supply systems, Fuchs argued (following Broshi) that around 20 BCE, the population of Jerusalem was about 38,500, while around 70 CE the population was about 82,500 (corresponding to a growth rate of about 1% per annum). Levine's estimate for around 70 CE was between 60,000 to 70,000, while Wilkinson's estimates for around 70 CE was about 75,000 persons. These are all in reasonably good agreement; to be conservative, Fuchs adopted Broshi's estimates.

Third, citing various sources, Fuchs estimated the birth rate to have been between 4% and 4.5% per year—corresponding to an average fertility rate of about 6 to 7 children per woman—and he estimated juvenile mortality to have been between 35% and 50%. Fuchs used the midranges in his computations, and a truncated Poisson distribution to model the number of children per woman estimating that approximately 132,200 Jerusalemites died in the period between 6 CE and 70 CE.

Of these, approximately 66,100 were male and 66,100 were female, counts which include infants, juveniles, adults, and non-Jews. Conservative estimates are that 5% of the population were non-Jews and that 42% of the deceased were juveniles, leaving 36,420 male and an approximately equal number of female deceased Jewish adults during this period.

Next, to afford a tomb-site and other costs associated with ossuary burial required some degree of affluence. As well, ossuaries bearing inscriptions evidence some degree of literacy on the part of the family involved. Literacy and affluence were no doubt correlated attributes, and Fuchs concluded, using a sequence of relatively conservative estimates, that at most 12% of the population satisfied these dual criteria. This led him to a "relevant population size" of around 4,370 males at most buried in inscribed ossuaries in the Jerusalem area during the relevant era. To place Fuchs' estimate in context, recall that the State of Israel collections (as itemized by Rahmani) contained only 233 ossuaries bearing inscriptions (with some being of women) and that in Ilan the names of 519 male persons were derived from ossuaries (with some only being fathers on mens' as well as on womens' ossuaries). Fuchs' estimates thus appear to be both reasonable and conservative.

Fuchs did not require nor did he estimate the number of ossuaries of females bearing inscriptions. Since among Ilan's ossuary sources 519 male and 193 female names were found, it appears that 27% of inscribed ossuaries bear female names—a male to female ratio of about 2.7 to 1. Of course, this underestimates the proportion of inscribed female ossuaries. While one could more accurately estimate this proportion by pursuing fine detail in Ilan we propose instead to use a crude estimate based on a ratio of 2 to 1, namely that 2,185 females were buried in inscribed ossuaries. This estimate appears adequate for our purposes and (conveniently) corresponds with the ratio found in the Talpiyot tomb.

Relative to questions of whether or not the Talpiyot tombsite could be that of the NT family, the data from that site must be viewed as the "best" of many trials. So far, about 100 tombsites have already been explored, but the mere existence of others that have not been must somehow also be accounted for. The Talpiyot site consists of 4 male and 2 female inscriptions. When divided into Fuchs' estimates for the total number of inscribed adult ossuaries, we obtain approximately 1,100; this appears to be an appropriate number of trials out of which the Talpiyot observation could be considered as being the "best."

**9. Inferential issues.** This section concerns whether or not statistical reasoning applies to this problem, and whether the available data permit meaningful analysis of an archeological find such as this. Remarks regarding the interpretation of "tail areas" are postponed to Section 14.

Several issues need to be addressed. First is the "fear-factor" connected with proposing an analysis on a controversial topic; it seems fair to say (and certainly in hindsight) that the intensity with which any analysis of this data set will be scrutinized constitutes an arguably unprecedented feature of this problem. Faced with this one may be tempted to adopt so highly conservative a stance that all evidence becomes masked. We side-step this and try to analyze the data as in an ordinary statistical problem; the resulting computations must then to be interpreted by each "consumer" for themselves. Second are "theological" considerations which if rigorously adhered to void any possibilities for analysis. The approach we adopt is to analyze the data from a purely "*historical*" *viewpoint*, by which we also mean that all persons referred to are assumed to have been real and subject to all considerations real persons are subject to. Third, there is the question of whether the available data bear adequately on the problem at hand. One could argue that the available onomastica cannot be authenticated (i.e., matched to the actual populations) and so on. We bypass such viewpoints and adopt the position that considerable and relevant data are available for the problem at hand.

Harder to dismiss is the role of "coincidence" [see Diaconis and Mosteller (1989)], the issue being that this data did not originate in a planned experiment; coincidences occur all the time, and their a priori probabilities can be extremely small, even though the probability is not small that *some* coincidence will happen to *someone, somewhere, sometime.* It could be argued that such data cannot be analyzed, or that extremely minute levels of "significance" are required to carry evidentiary value. A kind of "relativity" operates here toward which the analyst must adopt a stance. For our problem, to an "observer on the ground" in Jerusalem interested only in results from digs, these data originate in a standard way. It is tempting to argue that because this find concerns the most well-known family that ever lived it actually might exempt us—*purely on technical grounds*—from the limitations of coincidence. In any case, our analysis will be carried out from the vantage of the aforementioned "observer on the ground" in Jerusalem.

There are also certain subconscious and/or widely held misperceptions that "interfere" in our attempts to assess the evidence in these data. In particular, one needs to face the fact that it does seem extraordinary, at first, to contemplate that an ossuary that may have been intended for Jesus of Nazareth could ever possibly be found. The following *historical* point therefore needs to be made: Jesus was a Jew—a devout man who followed the letter and the spirit of the Jewish laws, as did other members of the NT family. Unless prevented by *force majeure*, the family (and followers) of Jesus would have certainly seen to a quick and proper burial in accordance with the Jewish ritual customs prevalent at the time. Roman authorities saw to Yeshua's crucifixion because they deemed it against their interests to allow a man proclaimed as being "King of the Jews" to live, and for the same reason would have certainly executed any son(s) of such a "King." But there is little reason for Roman authorities to have stood in the way of families of crucified persons from subsequently conducting proper burials, and there are in any case accounts of how release of the body was secured through the influence of Joseph of Arimathea. In fact, Joseph of Arimathea offered a burial site, in Jerusalem, for that purpose (as evidently the NT family did not yet have one of its own) and the single most likely eventuality, from a purely historical stance, is that the remains of Jesus were intended for interment in an ossuary—although possibly as much out of the sight and knowledge of Roman authorities as possible. Moving the remains to (say) Nazareth—a trek of some three or four days—may hardly have been feasible considering logistics at the time; indeed the Talpiyot location is among the many where one might reasonably expect such a tomb—if one existed—to be found.

Next, the ossuary inscribed "Yehuda son of Yeshua" plays an unsolicited role in the inference because at least this much is true: If this tombsite really were that of the NT family, then there did live a person named Yehuda whose father happened also to bear the name Yeshua. In that eventuality, the possibility arises that the two Yeshuas may have been the same person. It would not have been considered unusual for a Jewish man to have a child, and if that child was believed to be a target of the Romans, it would not have been unusual to try to protect it. However, other possibilities exist as well, with the time elapsed between the crucifixion and the destruction of Jerusalem allowing other scenarios to have played out. If, on the other hand, an ossuary inscribed "Yehuda son of Yeshua" may (for whatever reason) not be located in a NT family tombsite, then the Talpiyot site cannot be that of the NT family and the names found there must be purely coincidental. In our analysis, this ossuary will initially be "set aside," but we revisit this in Section 14.

Experimental design issues (as well as their absence) also play a role as there are several hypothetical scenarios under which our data could, *in principle*, have been collected. Furthermore, we do not know a priori whether or not a NT tomb site actually exists, the individuals who might have been within it, or the renditions of their names—considerations which each subtly affects the character of our inferences.

Conditioning and/or ancillarity, which are standard statistical practice, play an especially important role in our analysis. It seems reasonable, and perhaps even a practical necessity, in analyzing these data, to condition on the number of inscribed ossuaries found in this tomb, and also to condition on the fact that two were female and four were male. In some respects, these values carry little "information" relevant to the questions of interest here. We also condition on the fact that two of the inscribed male ossuaries are aligned in the generational sequence "C son of B" and "B son of A." The fact that there were a total of ten ossuaries in the tomb may or may not be viewed as ancillary, but not the ratio 6/10 of inscriptions, for that ratio carries information concerning the "literacy" of the family that owned the tomb. Likewise, the specific languages used on the inscriptions cannot be regarded as entirely ancillary because some information is available about the languages used by NT family members. Conditioning will thus play a significant role in our analysis, with even our "test statistics" permitted to depend on certain observed configurational aspects of the find.

A further inferential issue is that more than one reasonable analysis may be proposed (even by the same statistician) leading to somewhat differing "*p*-values." C. R. Rao recently referred (2007, at Cochin) to a 1992 Leiden Ph.D. thesis by Van den Berg which consisted of sending the identical data set to ten renowned statisticians, resulting in ten different analyses. Andrews and Feuerverger (2005) have argued that examining a collection of models allows the variations among their results to speak for the true inherent uncertainties without trivializing a problem.

As a final point, we mention that NT genealogical data is subject to considerable ambiguity, with names having frequently changed in form across sources, across time, and across translations. Care must therefore be exercised to assure that any proposed analysis is not influenced unduly by prior examination of the data, a principle well enough understood, but difficult to incorporate in practice.

**10. Our "a priori" hypotheses.**    In Sections 11 and 12 we develop a statistical approach based on "relevance" and "rareness," or "surprisingness," for addressing questions such as those raised by the Talpiyot site. Here—*on a best efforts basis*—we attempt to formulate a reasonable set of a priori alternative hypotheses. Our approach is strictly "*historical*" and with no claim made, of course, that the data has not been seen. We propose eight a priori hypotheses (APH) in all.

- **APH 1:** An ossuary intended for Jesus was likely to have been produced in the Jerusalem area. He was first laid to rest near the site of the crucifixion under the initiative of Joseph of Arimathea,[16] and it is unlikely that followers would have dishonoured the body in any way.
- **APH 2:** It is likely that one or more among the more affluent followers of this Messianic movement would have seen to a tombsite for the NT family and/or for some of its key leaders.
- **APH 3:** Inferring from biblical accounts, if there were a NT tombsite, no one who predeceased Jesus may be in it. One such *person* is Joseph, the father. (This does not preclude the *name* Joseph from occurring in the tomb.) Another such person is John the Baptist.[17]
- **APH 4:** No one who died after 70 CE may be found in such a tomb. Hence Simon and Yehuda will be excluded (although their names are not). This also excludes most—although not all—of the apostles, many of whom lived beyond 70 CE.
- **APH 5:** Closest relations, particularly closest blood relatives, are among those who might be expected to be in the tomb. Among those whose names are essentially known, are the mother Mary, brothers James and Joses, sisters[18] Mariam and Salome, and as a more remote possibility, a third sister Joanna. Potential blood relations or others *very* close to the family can also be identified from among those present at the burial ritual. This includes a Marya (referred to as the mother of James and Joses); it includes Mary Magdalene[19] whose presence at the burial ritual

---

[16] This NT account suggests, incidentally—and it is an important point for us—that the NT family did not yet have a tombsite of its own.

[17]According to Josephus, John the Baptist died [was beheaded] at Machaerus before Jesus. He was thus most likely buried at Qumran, or in the vicinity the Dead Sea.

[18]The likelihood of a sister being in a NT tomb depends in part upon whether or not she was married.

[19]Although Mary Magdalene is sometimes cited as a possible *candidate* for a "spouse" on the basis of her presence at the burial (confirmed in Mark 15 and Luke 8), and on the basis of later gnostic sources which refer to her as a companion of Jesus, our analysis does

is consistent across all gospel accounts; and it includes a Salome who might be a sister of Jesus.[20,21] The list of family intimates might also include a sister of Mary and/or possibly her spouse Cleopas (generally assumed to be the brother of Joseph).

- **APH 6:** The tomb might include close associates and/or others mentioned prominently or strategically in the NT (e.g., some apostles, especially if related through blood and/or marriage), close friends, and/or slightly more distant relations of the family. It would exclude anyone whose tomb has already been found elsewhere, or who is known to have lived and/or died elsewhere. A brief discussion of potential such persons is given below. The a priori probability of inclusion for individuals in this group is less than for those in APH 5, and their number would be related to the size of the tomb complex. Because the genealogy of the NT family is not known fully, such a tomb might also contain individuals whose names are unknown (or would not have occurred) to us.

- **APH 7:** It would be expected that if a NT tomb existed it might be unusual or distinctive in some way, reflecting the prominence or other characteristics of the family via some feature(s) of the site; exactly how, one cannot say. As the NT family does not appear to have been large it is plausible that their tombsite might also not be large.

- **APH 8:** There is no a priori hypothesis as to the number of ossuaries that might be found in such a tomb, as to their configuration, or as to the renditions of names that might appear on them,[22] but it might be expected that these ossuaries would in some respects be unusual, with some bearing distinctive or unusual inscriptions and/or ornamentation, and perhaps more detail in the rendering of names than typical.

Let us next consider, in a little further detail, the persons (or names) that might be viewed as candidates for inclusion under APH 6. Those present at the funeral have already been discussed. Among others mentioned prominently in the NT are individuals named Joanna and Suzana mentioned in Luke 24:10 as providers of financial support. The name Martha also appears in the NT as a close friend but she came from Bethany and would likely have

---

not assume this; it only assumes that she is on a "short list" of persons close enough to the family to be a candidate for inclusion in a NT tomb, an assumption which is by no means universally accepted.

[20]The brothers are not named as having been at the burial and most likely fled (as did the other apostles) for fear of their lives; none was present at the crucifixion.

[21]A woman called Martha (whose brother was Lazarus) may also have been present at the burial, however her ossuary is believed to have been found at Dominus Flevit.

[22]On the other hand, an ossuary inscribed "Shimon bar Yonah" found at Dominus Flevit and believed to correspond to one of the apostles helps us to infer what a NT inscription should look like.

been interred in her own family's tombsite there. As concerns the apostles—
many of whom are believed to have survived beyond 70 CE—there are no
substantive a priori reasons for any of them to be found in a NT family
tombsite—especially if it were a small one—unless related by blood to the
family; this would be the case if the apostle happened also to be a brother.

As evident from the discussion, the a priori candidates for a NT tombsite
are not unlimited. Of course, from this information, more than one plausible
a priori list can be constructed. However, we will work with different possible
lists as well as with different numbers of (and frequencies for) candidates.

We can now write down our a priori list of candidates for a NT tombsite. In
alphabetical order, for the women, this list includes, initially, the *persons*[23]

> Mariam,      Mary,      Mary Magdalene   and   Salome.

For the men, it includes

> James,      Jesus   and   Joses.

In expanded versions, the lists may include

> Cleopas,      Joanna   and   Martha,

although these persons are considered to be more remote possibilities. The
list of *persons* (but not necessarily *names*) that would *disqualify* the tombsite
as belonging to the NT family includes

> Joseph,      Simon   and   Yehuda,

as well as many rather specific and/or unusual *names*[24] thought not to be
associated with the NT family in any way. The consequences of not speci-
fying a disqualifier list more fully will be statistically conservative. Finally,
the list of names that do not disqualify the find, but that otherwise offer
little or no "evidentiary value" is lengthy; for our purposes, it will suffice for
this list to consist of all names other than those already included here.

Next, we need to deal with the fact that even if the ossuary for a candidate
on our lists were found, we have no way of knowing a priori by which *rendi-
tion* their name would appear. Our paradigm for measuring "surprisingness"
will allow us to handle this problem in an effective way, but will require an
a priori assignment of a measure of "surprisingness" to any name rendition
that might occur. It will be more convenient to deal with a reciprocal form
of "surprisingness"; this will be a measure of "relevance and rareness" which we will call the "RR value." "Relevance" will refer to membership in an a priori list of tomb candidate name renditions. The RR value of a datum, or of a subset of data, will often be the same as the frequency of occurrence of its "relevant" components under independent random sampling from the onomasticon, but there will be essential exceptions to this. (The complete definition is somewhat involved and will be detailed further below.) For the sake of definiteness, we define "surprisingness" as $-\log(\text{RR value})$, or alternately as $1/(\text{RR value})$.

---

[23]The four lists given here, are not lists of *names*, but of NT *persons;* here Mariam, Sa-
lome and Joanna refer to (possible) sisters, James, Joses, Simon and Yehudah to brothers,
etc.

[24]Certain specific *renditions*, even for *generic* names associated with the NT family,
could also be included in this disqualifier list.

The way in which we shall assign "RR values" to name renditions of NT persons on our a priori candidates lists is via *prespecified* nested classes of sets of name renditions in which the innermost class(es) represent the most "relevant" but "rarest" (i.e., the most specific but appropriate) renditions of that person's name, and the outermost classes include the less rare renditions still considered relevant for that person. These classes are compiled in conjunction with the totality of the information in Ilan (which includes the Talpiyot names). Collections of outermost sets of such nested classes may themselves constitute a part of a partition of the generic name category from which they derive, as may happen when the generic name applies to more than one NT individual. This occurs in particular with the generic name Mariam which here can refer to three different "intimates" of the NT family, and with the generic name Joseph which can refer to two such individuals. Nothing here is intended to prevent the same rendition category from applying to more than one person.

To now become specific, for Mary Magdalene, we initially allow a nested class of renditions consisting of the following three "appropriate" and decreasingly rare sets: (a) the set consisting only of the rendition Mariamenou [η] Mara; (b) the set consisting of all versions of Mariamne, including the one in (a); and (c) the set consisting of all Mariams,[25] including those in (b). Upon consulting Table 4, we observe that no (nonfictitious) rendition of Mariam appearing in category (b) and not also in category (a) occurs among sources restricted to ossuaries. Since only ossuary-based sources ultimately figure in the analysis, our categories (a) and (b) here actually become identical; we are thus left with only two nested rendition categories for Mary Magdalene. Now, each such renditions set will have an a priori RR value associated with it, and when an observed rendition of a relevant name is encountered, the RR value associated with it will be that of the rarest set to which it belongs. The specific measure of "rareness and relevance" associated with such a set will be defined below; typically it will be the relative

---

[25]It is possible that for Mary Magdalene only the renditions in (a) and (b) are relevant and that the remaining Mariams in (c) are not. However, the results of our analysis will not depend upon whether or not we include (c) here since it becomes included upon considering Mary.

frequency of that set within the onomasticon, but certain exceptions to this will be permitted.

Continuing in this way, for (the mother) Mary we allow two classes, namely (a) all versions of Marya; and (b) all Mariams,[26] including those in (a). For the (possible) sisters Mariam, Salome, and Joanna we have only the generic name sets for each since none is known by any rarer rendition; this applies to Martha as well.

For the men, we must be mindful that the name of the father on the generational ossuary plays a different role than the other male names. In any case, for Jesus,[27] as well as for James and Cleopas, we have again basically only their generic name categories, while for Joses we have (a) all renditions consistent with Joses (as at the bottom of Table 5); and also (b) the generic Joseph set. As for the father on the generational ossuary two additional persons are relevant for us, one being Joseph, the father of Jesus, and the other being Jacob, the father of this Joseph—but the latter relevant primarily because he is also the possible father of Cleopas, and relevant only if the generational ossuary were to read "Cleopas son of Jacob." For Joseph and Jacob we again have only their generic name classes associated with them, as neither appears to have been known by rarer renditions. The RR values assigned to these renditions will be context-dependent owing to configurational considerations induced by the presence of the generational ossuary.

**11. A proposed method for analysis: preamble.**    We turn now to develop our approach for the inference problem at hand. Because application of a classical hypothesis testing framework in the present context is not straightforward, we consider an approach centering generally on the "surprisingness" of observations, and of how frequently—under a random sampling protocol from the onomasticon—a cluster of observations of equal or greater "surprisingness" would arise. The idea is to try to circumvent specifying aspects of an alternative hypothesis "inessential" to the problem. Broadly put, "surprisingness" is related (inversely) to 'relevance and rareness" in observations (referred to as "RR" values), with "relevance" referring generally to association of the data with what might be expected to occur in a tomb of the NT family, and "rareness" connected with, but not identical to, a relative frequency associated with those data.

---

[26]Conceivably, one could argue here to omit the broader class (b) and allow only (a). However, owing to the presence on our list of a sister whose name might be Mariam, this decision again is inconsequential.

[27]Strictly speaking, Jesus was known *only* by the version Yeshua of the generic Yehoshua. However, the full and formal Yehoshua is *never* used in Second Temple documentary texts [T. Ilan, private communication] and for this reason we allow only the generic name category here.

An approach based on "surprisingness"—or "RR" value—possesses some useful features: First, it provides a more "natural" method for specifying relative probabilities for clusters of names under the alternative. It also permits us to deal effectively with the fact that names of "relevant" persons can present in more or less rare renditions; such renditions may be nested and different "RR" values assigned to them. Furthermore, it leads us, in a natural way, to recognize that under the *alternative* hypothesis, the probabilities associated with any given set of names are not invariant under configurational rearrangements of those names; it also provides an intuitively natural way to encode subtle features of the probability structure arising out of the complex of family interrelationships. The method is also useful in helping distinguish between those aspects of the alternative that are of an a priori nature from those that are a posteriori; in particular, it allows us to more easily recognize that the test procedures can themselves be allowed to depend upon certain aspects of the observed tomb *configuration*. Last, but not least, the method affords us the convenience to ignore names whose evidentiary values are regarded as negligible, even though many such names would be viewed as not inconsistent with a NT tombsite. Such features make the method easier to implement than a carefully crafted likelihood ratio test which requires a precise specification of an $H_1$-probability structure. Any seemingly "incorrect" specifications of the alternative hypothesis will only result in some modest losses of power.[28] We see it as not disadvantageous to make that sacrifice, viewing it as partial payment toward any inadvertent *post hoc* indebtedness in the inference.

Returning to our discussion on measuring surprisingness, if we were to mirror a standard hypothesis testing setup, $H_0$ might be the assertion that the observed configuration of names arose by purely random draws from the onomasticon; an alternative "$H_1$" would be an opposite of $H_0$ relevant to the "NT hypothesis" that the tombsite is that of the NT family. A "sample space" would consist of all possible drawings from the onomasticon, subject to the conditioning of there being two women, and four men, two of whom are in father-son generational alignment. Some modest "realism" restrictions on points in the sample space may also be required; specifically, within a small tomb, the exact name renditions of deceased persons ought to differ. We next need to order the points in the sample space "along an $H_0$—$H_1$ continuum." (This occurs naturally in the classical setup once $H_1$ is specified fully.) It cannot be entirely unambiguous as to how such an ordering should be defined; loosely put, we want to order points on the basis of how "convincingly" they reflect what one might expect to find in a NT family tombsite. Among name clusters not inconsistent with "$H_1$" this ordering could be on the basis of the probability, under purely random sampling, of prespecified aspects of the cluster that most convincingly "allude" to "$H_1$." Thus, for example, the presence of a name such as Matya should not disqualify a cluster (since it is not inconsistent with the genealogy) although its probability contribution might be discounted (e.g., set to unity)—for we have, after all, no idea who this may be—while a more "rare but relevant" name such as Yoseh should have its probability accounted for in the computation. The "probability," or "RR value," resulting from a computation of this type (with the familial and other adjustments to be discussed below) will be used to measure "surprise"; smaller RR values $\Rightarrow$ greater surprise. A "tail area" for assessing "evidence" against $H_0$ would then be based on the overall probability, under the $H_0$-sampling, of the set of points in the sample space whose RR values are less than or equal to that of the observed outcome (i.e., which are as or more "surprising"). If this "tail area" is sufficiently small, we may then consider to invoke the standard logic and conclude that either we have witnessed an event of rare chance, or the null hypothesis must be untrue. We are being cautious not to use the term "$p$-value" here; a more careful discussion of the interpretation of a small tail area will be undertaken in Section 14. For an appropriate definition of "surprise"—which must be specified a priori—a key computational question then becomes: *What is the probability that a (permissible) random sample of two female and three[29] male ossuaries, configured as at Talpiyot, contains a cluster of names which (relative to this $H_0$ and "$H_1$" setup) is as or more "surprising" than the cluster found?*

---

[28]This occurs because the presence in the sample of nonrelevant names, and the absence of relevant ones is not fully "optimized" for, although such "mathematical" optimality here is more apparent than real.

It is perhaps worth remarking that if we proceeded classically and carried out a LR test on the basis of a priori hypotheses such as APH 1–APH 8, then if a Wald's $\chi^2$ type of approximation were applicable we would need the probabilities (under $H_0$ and $H_1$) only for the observed data point, that is, only for the names and configuration observed. But whether such a test is carried out exactly via enumeration, or only approximately via a Wald's approximation, the out-of-sample names will matter only as to their number and their probabilities, the actual names themselves will not matter; and in turn, their number and their probabilities are required only for determining the distribution of the LR test statistic under $H_0$. The imprecision in this assertion pertains primarily to matters concerning the tomb configuration and familial interrelationships among the names. But if one already includes within the alternative those names that are "configurationally active," thereby accounting for their contribution to the overall "$H_1$" probability structure, the inclusion of additional names becomes essentially straightforward, and our assertion then holds more precisely. This opens the possibility that seemingly quite different versions of "$H_1$" could lead to essentially similar test results. As long as two versions of "$H_1$" were not particularly opposed to any of the names in-sample, but otherwise had (possibly quite) different sets of out-of-sample candidates, although approximately the same in number and with comparable "$H_1$" probabilities, then the results of the tests should be similar. The robustness of any procedure—specifically to the "$H_1$" specification—could then presumably be checked by allowing for different numbers of out-of-sample names, and different probabilities for them, with the actual names themselves not mattering. Robustness to *moderate* variations in the "$H_1$"-probabilities of in-sample (as well as out-of-sample) names could presumably also be checked in this way, although the same cannot be said for *contentious* "$H_1$"-disagreements concerning any of the in-sample names.[30] It is to be understood, throughout our discussions, that all versions of "$H_1$" require a broad category of "Other" for all of the essentially "uninformative" names that could occur but are not otherwise considered to be inconsistent with "$H_1$."

---

[29]The Yehuda ossuary is initially being excluded in our analysis.

One final point arises from the fact that even if a "person" in the tomb is on our "$H_1$" a priori list, we do not know what rendering of their name will occur, and in particular how "relevant and rare" that rendering will be. If a name version is rare, this would be evident from Ilan's lexicon. However *rare names are not rare* and there may well be more than one possible rare rendering for any particular individual. In the end, however, the ossuary of such an individual would have been rendered in (at most) one such way. Hence the "rare names are not rare" concern does not apply so much to any particular individual's many potential rare renderings, as it does to the case that too many *persons*, each having rare name forms, are all considered to be likely candidates under "$H_1$." Accounting for this requires that we carry out the analysis allowing for more rather than fewer possible candidates having rare names.

**12. A proposed method for analysis: the RR method.** With the background of the previous section behind us, we may now describe in further detail our proposed paradigm based on "surprisingness."

---

[30]In the present context, one of the more contentious "$H_1$"-disagreements centers around the inclusion of Mary Magdalene as an "$H_1$"-candidate. We note that—*on purely technical grounds*—this contentiousness makes her an ideal "$H_1$"-candidate for *some* "hypothesis test." In any case, it may be that the contentiousness of Mary Magdalene as an "$H_1$"-candidate has arisen because some interpret this as intending to imply that she was a spouse to Jesus although no such assumption is made here. Sensitivities to this issue appear to have been heightened due to a recent fictional account; see Ehrman (2004). Another source for this contentiousness possibly arises from the tradition that regards Mary Magdalene as a "sinner"; the earliest historical accounts, however, do not corroborate that view.

Because our inference is conditional on the observed configuration, our procedure may depend on that aspect of the observed data (although not on any other). For the sample of the two women, we (initially) consider the case that "$H_1$" allows selection from a list of persons which consists of Mary Magdalene, Mary, Mariam, Salome, and "Other," together with their corresponding name rendition classes as defined in Section 10. In numerical experimentation, this list may be reduced, and/or augmented by Joanna, Martha, Woman(1), Woman(2), etc., where "Woman(i)" is considered to be a "relevant" (out-of sample) person whose name is left unstated.[31] This list is intended to reflect APH 1–APH 8. The category "Other" groups together all other female names, in particular those considered not to be informative. Any selection of a female name from the above list is "relevant," except for "Other." The "RR" value for each of the names on this list is typically, but not invariably, the probability of the rarest rendition category (among our pre-defined categories) of the observed version of that name under random sampling from the onomasticon; the name category "Other" is discounted by being assigned an RR value of 1. The RR measure (relevance and rareness) for a set of two womens' names is defined as the product of their individual RR values. The sampling of these womens' names is carried out by drawing independently from the onomasticon, except that we do not allow any name *rendition* to occur twice.

Turning next to the men, the list of persons under "$H_1$" is taken initially to consist of Joseph (as a father), Jesus, Joses, James, and "Other," to be augmented in numerical experimentation by Cleopas, Male(1), Male(2), etc., with corresponding rendition classes again as given in Section 10. The RR values for each of these name renditions are computed from their onomasticon frequencies, except for the uninformative category "Other" which is assigned an RR value of 1 and is otherwise treated as before. With these conventions, male names are selected under random sampling from the onomasticon, and assigned to the two singleton male ossuary slots, and the two slots on the generational ossuary. This random sampling for the men is restricted by realism requirements to ensure that "no man dies twice," and that a father and son have different names. The RR value (relevance and rareness) for the sample of the four male names is then defined as the product of the RR values of the individual names, except for adjustments deriving from NT familial relations detailed below. The RR value for the combined male and female sample has yet to be defined; for the moment

---

[31] The actual names will be unimportant except for those in-sample; only the number of such name categories and their probabilities or RR values will actually matter. This will also be the case for some of the configurational aspects operative in the case of the men.

may we take it to be the product of the RR value for the females and the RR value for the males.

The exceptions to the $H_0$-sampling may be summarized as follows. For the women, we do not permit any name rendition to occur twice. Likewise for the men, for any configured set of four male names we do not permit the name renditions of the two singleton males to be identical (unless both are "Other") as one would (again) not expect two inscribed ossuaries to have been left indistinguishable in a small tombsite. Furthermore, we do not permit the father and son name renditions to be the same (unless both are "Other"). And finally, we do not permit the name rendition of the son to also be that of one of a singleton male (unless "Other"), the idea being again that "a person cannot die twice."

We next indicate the nature of some of the definitional adjustments to the RR value imposed by "$H_1$." As it happens, these involve only the names for the males, and often involve the father in the generational ossuary. Typical among such restrictions and adjustments are the following. If the father is "Other," then the RR value for the generational ossuary is set to 1 regardless of the son's name, for we then do not know who that son may have been and therefore discount it. Next, the father of that pairing is not permitted to be Yeshua[32]; in that case we set the RR value for the pairing to 1, or even to $\infty$, the net effect being about the same. If Yosef is the father and the son is "Other," and if that Yosef cannot be the biblical brother by virtue of there also being a Yosef in the tomb, the RR for the generational ossuary is set to 1, since we again do not know who this Yosef is. These considerations are far from complete; a complete set of restrictions and adjustments of this type will be detailed in the following section.

Finally, the RR value for the entire sample is defined as the product of the RR value for the females and the RR value for the males but possibly with exceptions of the following type: We may consider requiring the name Yeshua to appear as either the son in the generational pairing or as one of the singleton males; the idea here is that nothing beats the "surprisingness" of the *ne plus ultra* name Yeshua—appearing in a consistent manner—in a tombsite being gauged for having belonged to the NT family. Nevertheless, the inferences do need to be checked for robustness to requirements of this nature. As long as the definition of "surprise" (or "RR" value) is specified fully and a priori, the resulting approximate "tail area" will essentially be "valid"; all that is still required would be to determine the distribution of the "RR" values under the null hypothesis.

---

[32]Having one Yeshua in the tomb as a father is "problematical" enough; a second is not being permitted.

**13. A statistical analysis.**   In this section we summarize a statistical analysis of the Talpiyot tomb data based on the paradigm developed in the previous two sections. Our analysis, however, is predicated upon a particular set of assumptions. Statistical analysis often follows from factual direction by subject matter expertise—in this instance from specialists in the history of early Christianity, in ancient scripts and carvings, and so on. The assumptions A.1–A.9 under which we carried out our analysis[33] are by no means universally agreed upon. Furthermore, the failure of any one of them can be expected to impact significantly upon the results of the analysis. We begin by itemizing these nine assumptions.

- **A.1:** We assume the "physical facts" to be correct: that the Talpiyot burial cave was found and provenanced properly, that it had remained essentially undisturbed since antiquity, and that no ossuaries were moved into or out of the tomb between the time the burials took place and the time in 1980 when the tomb was excavated.

- **A.2:** We assume that if any ossuaries bearing inscriptions were removed from the tomb they were removed *haphazardly* and with no intent to mislead "in the direction" of "$H_1$"—that is, without regard to inscriptions that may have been inconsistent with "$H_1$."

- **A.3:** We assume that the historical and genealogical information relied upon here is adequately accurate. In particular, we assume that the most appropriate rendition of the name for the mother is Marya, for the father is either Yehosef or Yosef, and that those for the siblings are as given in the NT, with the second brother's (Yoseh's) most appropriate name rendition being as in Mark 6:3 of the NT.

- **A.4:** We assume that the ossuary inscribed "Yehuda son of Yeshua" can be explained and may be disregarded in our analysis. (We shall revisit this point in Section 14.)

- **A.5:** We assume the approximate validity of the demographic estimates for Jerusalem, in particular for the number of Jewish adults deceased within the relevant time spans, for the number of ossuary burials that took place, and for their inscription rates.

- **A.6:** We assume that the lexicon of Ilan (2002) provides a sample of names of persons from the relevant era sufficiently representative for our purposes, and that our implementation for their frequencies is appropriate.

- **A.7:** We assume that the full inscription Mariamenou [$\eta$] Mara refers to a single individual and represents the most appropriate specific appellation for Mary Magdalene from among those known; we further assume that this inscription is sufficiently distinctive that it could only have applied to very few and/or very particular individuals within the generic Mariam name category. Our specific implementation of this assumption will be of the type to assume that essentially at most one out of every 74 Mariams could legitimately have been rendered in this way, and that Mary Magdalene was among those who could.

---

[33]These assumptions were proposed by S. Jacobovici, except for A.6 and A.9 which are due to the author.

- **A.8:** We assume that the inscription of the father "Yehosef" on the "Yeshua ossuary" and the inscription "Yoseh" on that individual's ossuary were meant to distinguish among two different persons.
- **A.9:** We assume that on a *time cross-sectional* basis, the assignment of names is adequately approximated by independent sampling; thus, for instance, that fathers called Yehosef would name a son Yeshua with about the same incidence as occurs in the general population, and so on. (See also Section 14.)

We turn now to our analysis, stressing again that it is predicated upon *all* of the hypotheses APH 1–APH 8 and the assumptions A.1–A.9. We compare "surprisingness" (or rather "RR" values) for Talpiyot-like configurations of names, when sampled randomly from Ilan's onomasticon, with the corresponding values for the arrangement actually observed; these computations were based on complete enumeration over the onomasticon.

Our *baseline computation* involves sampling from the womens' name rendition categories

$$\text{MM,} \qquad \text{Marya,} \qquad \text{Mariam,} \qquad \text{Salome} \quad \text{and} \quad \text{Other,}$$

with relative frequencies

$$\frac{74 \times (1/44)}{317} = \frac{1.68}{317}, \qquad \frac{74 \times (13/44)}{317} = \frac{21.86}{317}, \qquad \frac{(74 - 1.68 - 21.86)}{317},$$

$$\frac{61}{317} \quad \text{and} \quad \frac{317 - 74 - 61}{317},$$

and assigning to these renditions the "RR" values

$$\frac{1.68}{317}, \qquad \frac{21.86}{317}, \qquad \frac{74}{317}, \qquad \frac{61}{317} \quad \text{and} \quad 1,$$

respectively; here MM stands for "Mariamenou [$\eta$] Mara" (or equivalently for our data, just Mariamne). The frequencies for MM and Marya were discussed in Section 7; the frequency for Mariam is based on the complement in the set of generic Mariams[34] after the MMs and the Maryas are removed. The RR values assigned to the name categories are the same as their corresponding assigned frequencies, but with several exceptions: The RR value for a Mariam who is not an MM or a Marya, is based on the frequency of the entire generic class; the rationale for this is that this is now a very common rendition of a very common name, and while it is consistent with the NT genealogy, it carries reduced evidentiary value. Also, the name category "Other" is assigned an RR value of 1; higher RR values still could be assigned to any womens' names thought to invalidate the find although we did not implement such an invalidation set—the impact of this being, of course, conservative.

---

[34] "Mariam" is being used in two senses here: as the generic name category, and as the "other" Mariams after the specialized ones are removed. This will also occur with the name Joseph. The intended meanings should be clear from the context.

The mens' name rendition categories for our *baseline computation* are

$$\text{Yosef}, \qquad \text{Yeshua}, \qquad \text{Joses}, \qquad \text{James and Other},$$

with relative frequencies

$$\frac{(221 - 33.63)}{2509}, \qquad \frac{101}{2509}, \qquad \frac{221 \times (7/46)}{2509} = \frac{33.63}{2509}, \qquad \frac{43}{2509}$$

and

$$\frac{2509 - 221 - 101 - 43}{2509},$$

and RR values

$$\frac{221}{2509}, \qquad \frac{101}{2509}, \qquad \frac{33.63}{2509}, \qquad \frac{43}{2509} \quad \text{and} \quad 1,$$

respectively. The category Other is again assigned an RR value of 1. The frequency (as well as the RR value) for Joses was discussed in Section 7. The RR value for Yosef is based (initially at least), on the full generic Joseph count—again on the grounds that it is now a most ordinary rendition, although for the renditions of Yosef the situation will actually be more involved since they could refer to either the brother or to the father; we shall need to revisit such issues below.

If, in numerical experimentation, any of our baseline name renditions are removed from our a priori lists, the adjustments required to the frequencies and RR values of the remaining ones are the natural ones. And if any names such as Joanna, Martha, Cleopas are added to that list, the frequencies and RR values associated with them will be based on Ilan's (nonfictitious persons) counts, namely 12/317, 21/317, 7/2509 (and so on), respectively. Updates to the frequencies for the categories of "Other" women and/or "Other" men are also the obvious ones.

To further appreciate the nature of the complications that may arise consider, for example, finding a Cleopas son of James ossuary. Should such a James be viewed as being the biblical brother with a hitherto unknown son? Or should this James be viewed as being the biblical grandfather? We are obliged to establish rules for differentiating among such possibilities.

The reader will hardly fail to notice—as our definition of "surprisingness" and "RR" value takes shape—the many judgement calls involved in their definition. Our choices are meant to mirror the intent that RR should essentially measure the probability contribution only for those aspects of the find that are considered *relevant* and *knowable* for the NT family; however what is important is that these judgement calls all be of an a priori nature and this we are attempting to do on a best efforts basis. Experimentation appears to confirm that "sensible" variations in the definitions do not make a great difference to the results of our computations—as long as one is operating within the same set of a priori hypotheses and assumptions, namely APH 1–APH 8 and A.1–A.9.

In addition to the "realism"-based sampling restrictions outlined in the last section, the computations in our *baseline* case involve a series of 14 configuration-related familial adjustments to the RR values whose interactions with each other can be a bit complicated. These were devised on the basis of what is believed known of the genealogy of the NT family and of our relative expectations of how one may have thought such names might or ought to be configured in a NT tomb. The parameters proposed below were all selected on the basis of appearing to be reasonable a priori choices, but the sensitivity of the computations to these choices was nevertheless checked to gauge their influence. For the *baseline case*, we now itemize the complete set of adjustments to the RR values as implemented in our "R" computing code [Feuerverger (2008)]:

- If the father is Yeshua, the RR value for the generational ossuary is set to 1.
- If the father is Other, the son's RR value does not count (i.e., is set to 1).
- If the father also appears as one of the singletons, his name is not counted twice toward the RR value.
- If the two singleton males are Yosef and Yoseh,[35] then under "$H_1$" we do not know who Yosef is and therefore set his RR value to 1.
- If Yoseh is the father then the RR value for the son is set to 1 since the biblical brother Yoseh did not have a son whose name we know. However, since it was not uncommon for sons to be named after close blood relatives we shall allow the particular names Yeshua, Yosef, James, and Cleopas for the son,[36] but in those cases we discount the RR value for those son's names by multiplying by 5.

---

[35]Note that the case where Yoseh is the son and Yosef is a singleton will get handled (*q.v.*) by the fact that if Yoseh is the son of anyone other than Yosef then he cannot be the biblical brother. The reverse case where Yosef is the son and Yoseh is a singleton will get handled (*q.v.*) by the fact that Yosef will then be an unknown person.

[36]These four names correspond to persons believed to have died prior to the year 70 CE.

- Likewise, if Cleopas is the father then the RR value for the son is set to 1, however, we shall allow the particular names Yosef, James, and Yosa for the son but in those cases we discount the RR value for those son's names by multiplying by 5.
- If Yoseh is the father, and a Yosef appears as a singleton, then we do not know who that Yosef is (even though this name is not considered to be invalidating) and so we assign to that Yosef an RR value of 1.

In respect of the next four points (with Yosef being the father in each), we bear in mind that the name Yosef can refer to either the biblical brother or to the biblical father, unless Yoseh is the name of the son or a singleton male, in which case Yosef can only refer to the biblical father or to someone we don't know; we must therefore make RR value adjustments to account for the resulting scenarios:

- If Yosef is the father but is not also a singleton male, and if Yoseh is either the son or a singleton—thereby ruling out that Yosef is referring to the biblical brother—then the RR value for the generational ossuary is set to 1, unless the son is either Yeshua, Yoseh or James, in which case the generational ossuary receives its "full" RR value.
- If the father is Yosef and is not also a singleton male, and if a Yoseh does not also appear in the tomb—thereby making it possible that Yosef refers to either the biblical father, the biblical brother, or to someone we don't know—then the RR value for the generational ossuary is set to 1, unless the son is either Yeshua or James, in which case the generational ossuary receives its "full" RR value.
- If the father is Yosef and he is also a singleton male, and a Yoseh does not appear in the tomb then he can only refer to the biblical brother or to someone we don't know. In either case we do not know the name of the son. For our baseline case we allow the son to be either Yeshua or James but multiply that son's RR value by 5, and apply the usual RR value for the name Yosef.
- If the father is Yosef, and Cleopas is the son, and if a Yoseh is nowhere in the tomb, then regardless of whether or not Yosef is also a singleton, we treat him as referring to the biblical brother. The RR value for the generational ossuary is then the product of the RR values for Yosef and Cleopas except multiplied by 5 since that son's name was not known.

In respect of the next two points (James being the father in both), we bear in mind that the name James can refer to either the biblical brother or to the biblical father of Yosef and Cleopas; we must therefore make RR value adjustments to account for the resulting scenarios:

- If James is the father and is also one of the singletons, then under "$H_1$" he can only refer to the biblical brother or to someone we don't know and cannot refer to the biblical grandfather Jacob. In this case we permit the son to be either Yoseh, Yeshua, or Yosef, or even Cleopas, but we multiply the son's RR value by 5.

- If James is the father and not also one of the singletons, then he can be referring to either the biblical grandfather or to the biblical brother. In that case, if Cleopas is the son the generational ossuary is given its full RR value, but if the son is Yoseh, Yosef or Yeshua, the son's rarity is multiplied by 5.

And one final adjustment:

- If Yeshua is the son, and Yosef is the father, then in the baseline case we apply a "bonus" factor to this "prize" case by dividing the RR value by 1.2.

In numerical experimentation, the "downweighting" factor of 5 for "unknown sons" was varied and we also could entirely disallow RR contributions for the names of such sons. We also could omit the 1.2 bonus factor for the Jesus son of Joseph combination. Further, we could also require that a Yeshua must appear in the tomb before it could be considered to be as "surprising" as that at Talpiyot. Experimentation confirms, however, that the results of the computations are not unduly influenced by modest variations in such specifications for the definition of the RR values as long as such rules are selected in a generally reasonable way.

We turn finally to the results of our computations which are based on exact enumeration over Ilan's onomasticon. There are, firstly, a total of $317^2 \times 2509^4 = 3.982 \times 10^{18}$ possible samples (of persons) that can be drawn from the onomasticon (if order is allowed to matter); of these, $3.608 \times 10^{18}$ pass our "reality" requirements—that is, approximately 90.6% of drawn samples are "valid." For the Talpiyot tombsite, the RR values are computed as

$$\frac{74 \times (1/44)}{317} \times \frac{74 \times (13/44)}{317}$$

for the women,

$$\frac{221 \times (7/46)}{2509} \times 1$$

for the singleton men, and

$$\frac{101}{2509} \times \frac{221}{2509} \Big/ 1.2$$

for the generational ossuary, with the RR value for the overall find then being the product ($1.451 \times 10^{-8}$) of these three RR values; this computation takes into account all of our baseline rules including the 1.2 bonus factor for the prized Jesus son of Joseph pairing. Next, for our baseline context, we found that $1.981 \times 10^{12}$ of the "valid" samples have an RR value less than or equal to that of the Talpiyot tomb—that is, are considered to be as or more "surprising" than the Talpiyot find; the proportion of these is $5.491 \times 10^{-7}$, or about $1/1,821,000$. Multiplying this proportion by 1,100, that is, by the estimated maximum number of Talpiyot-like tombsites that can be formed from all inscribed ossuaries that had been produced in that region and in that era—gives 0.0006041, or about $1/1,655$. The interpretation of such a "tail area" is discussed in Section 14.

One intuitive explanation for this (baseline) result is as follows. The names of the four males can be arranged in 12 different configurations—4 choices for father, then 3 for son, the other two being singletons whose order does not matter. In Talpiyot the 4 male names which occur there were found in their unique "best" configuration. Loosely put, this contributes a factor of about $1/12$ to the tail probability. When combined with the "rareness and relevance" of the Mariamenou inscription these largely counteract that we are looking at the best of 1,100 tombsites. The remaining names are not equally rare but they are nevertheless relevant ones and random sampling over the onomasticon does not beat them too easily, particularly when NT familial relationships are properly accounted for.

We next examine the sensitivity of this computation to the various parameter choices, restrictions, candidate lists, and so on, underlying the baseline case. (We do not, however, deviate here from any of the assumptions A.1–A.6.) The questions at issue here concern how far we can push the "$H_1$" specification before the results become meaningless. This "stress testing" work involves: (1) Adding additional candidate names to "$H_1$," and/or removing names; (2) Changing the probabilities or RR values for names in "$H_1$"; (3) Changing the numerical values of parameters; (4) Adding or dropping various "$H_1$" restrictions and/or configurational bonuses; and (5) Combinations of the above. To prevent this high-dimensional task from becoming unwieldy, we carry out such steps one at a time, as well as in judicious combinations.

The following tail areas are obtained under the indicated "single condition" changes from the baseline case:

- Require that Yeshua be in the tomb before it can be considered to be more surprising than that at Talpiyot: 0.000552.
- Remove the bonus factor of 1.2 for the Yeshua/Yehosef generational pairing: 0.000726.
- Reduce the rarity adjustment factor (of 5) for unknown sons by half: 0.000696.
- Double the rarity adjustment factor for unknown sons: 0.000604.
- Do not count unknown sons (set their RR value to 1): 0.000597.
- Remove Salome: 0.000367.

- Add Joanna: 0.00111.
- Add Martha: 0.00103.
- Add Cleopas: 0.00267 [worst case[37]].
- Reduce the frequency and RR-value for MM by half: 0.000181.
- Double the frequency and RR-value for MM: 0.000953.
- Reduce the frequency and RR-value for Yoseh by half: 0.000323.
- Double the frequency and RR-value for Yoseh: 0.00131.
- Allow the father on the generational ossuary to be named Yeshua: 0.000697.

The following results are obtained under the indicated "multiple condition" changes from the baseline case:

- Add Joanna and Martha: 0.00159.
- Add Joanna and Cleopas: 0.00463.
- Add Martha and Cleopas: 0.00429.
- Add Joanna, Martha and Cleopas: 0.00669 [worst case].
- Double the frequency and RR-values for MM and Yoseh: 0.00220.

In the next group of results, Joanna, Martha, and Cleopas are all included, this being the "worst" of the cases computed above.

- Remove bonus factor for the Yeshua/Yehosef generational pair: 0.00752 [worst case].
- Require that Yeshua be in the tomb before it can be considered to be more surprising than Talpiyot: 0.00380.
- Remove bonus factor for Yeshua/Yehosef generational pair but require that Yeshua be in the tomb: 0.00415.

In the next group of results, Joanna, Martha, and Cleopas are all included, and no bonus factor is used for the Yeshua/Yosef pairing; this is again the "worst" of the cases considered above.

- Do not allow the RR value for unknown sons to count: 0.00635.
- Reduce the rarity adjustment factor (of 5) for unknown sons by half: 0.00871 [worst case].
- Double the rarity adjustment factor for unknown sons: 0.00678.

In the next group of results, Joanna, Martha, and Cleopas are all included, no bonus factor is used for the Yeshua/Yosef pairing and the RR adjustment factor for unknown sons is reduced by half. (This is the "worst" of the cases considered above.)

---

[37]Adding Cleopas results in the greatest deterioration in "tail area" among "single condition" changes. Here, as well as in each block of results below, we indicate the "worst case" within the block. Shortly, we pursue "steepest ascent" based on such "worst case" results.

- Reduce MM (frequency and) RR-value by half: 0.00410.
- Double MM RR-value: 0.0193.
- Reduce Yoseh RR-value by half: 0.00414.
- Double Yoseh RR-value: 0.0173.
- Double MM and Yoseh RR-values: 0.0353 [worst case].

In the next group of results, Joanna and Cleopas are included, but Martha is excluded; no bonus factor is used for the Yeshua/Yosef pairing, and the RR adjustment factor for unknown sons is reduced by half.

- For the case just stated: 0.00594.
- Reduce MM (frequency and) RR-value by half: 0.00274.
- Double MM RR-value: 0.0132.
- Reduce Yoseh RR-value by half: 0.00281.
- Double Yoseh RR-value: 0.0116.
- Double MM and Yoseh RR-values: 0.0238 [worst case].

In our last group of results, Joanna and Cleopas are included, but Martha is excluded; no bonus factor is used for the Yeshua/Yosef pairing, and unknown sons are not counted toward the RR value.

- For the case just stated: 0.00423.
- Reduce MM RR-value by half: 0.00199.
- Double MM RR-value: 0.00944.
- Reduce Yoseh RR-value by half: 0.00190.
- Double Yoseh RR-value: 0.00836.
- Double MM and Yoseh RR-values: 0.0169 [worst case].

**14. Discussion and concluding remarks.** We begin with some remarks on our computations. In some respects, the results are driven by the conditioning on the observed configuration of the inscribed ossuaries in the tomb, and their number is fortuitously close to being "optimal" for "allowing detection." With more inscriptions the combinatorial growth of possibilities dilutes power and with fewer inscriptions the premium on "rareness" diminishes. (Fortuitous "relevant" rarenesses among the renditions which occurred also play a critical role.) However, even with this seemingly ideal number of inscribed ossuaries our "tail areas" become "not significant" if the set of a priori candidates for a NT tombsite and their sets of name renditions (rare ones, in particular) become too large. This also occurs if these lists exclude certain in-sample names and renditions, in particular the rare (and controversial) "MM."

A number of simplifications were used to bound computational labour. We have, first, not implemented a list of names which invalidate a find. However, doing so would only invalidate some of the samples under $H_0$ hence further reducing our "tail areas" since the Talpiyot site contains no such names;

therefore the effect of that simplification is conservative. In fact, even within
the generic names among our candidates, there occur renditions for them
that also belong on our list of invalid names, or should at least be treated as
"Other" so far as their contribution to RR value is concerned. The effects of
our not having done so are again conservative since (1) the frequencies for
the relevant names are then higher than they really should be, (2) because
some of these renditions do not then wind up on an "invalid candidates"
list, and (3) because these renditions are wrongly assigned "legitimate" RR
values in cases when they should have been treated as "Other." A second
labour-saving approximation involved not concerning ourselves unduly with
the possibility of drawing identical name renditions (for the two women, or
the two singleton men, or the father and son) when those names arose from
the "Other" names categories; needless to say this should hardly impact on
the results.

Certain additional items of "evidence" or "data" that may carry "infor-
mation" relevant (in varying degrees) to our problem have not been incorpo-
rated into our analysis because such observations do not typically correspond
to a priori hypotheses; the question of if, and precisely how, such information
can be quantified in a *formal* statistical analysis is therefore problematical.
The items of this type of which we are aware are: (1) The untypical carving
of the circle and upward pointing gable on the entrance wall of the tomb; (2)
The rightward leaning "cross" at the head of the Yeshua ossuary inscription
which might be thought more distinctive than a mason's mark (although
its meaning, if any, is not known); (3) The proximity of the tombsite to
the Temple; (4) The unusually high proportion (6/10) of ossuaries bearing
inscriptions; (5) The languages used on the inscriptions, and in particular
the use of Greek script on Ossuary #1; (6) The fact that these ossuaries
are all of adult size; (7) Purported mitochondrial DNA evidence suggest-
ing that Yeshua and Mariamenou were not "maternally" related; (8) The
alignment of the three names Yehosef, Yeshua, and Yehuda which appear
on the two generationally sequenced father-son ossuaries ("A son of B son
of C") being the *only* one among the six possible arrangements for those
names that does not *immediately* invalidate the find; (9) Purported electron
microscopy tests which suggest that the spectral element signature of the
patina of the James ossuary matches to the Talpiyot tomb; and finally, (10)
The relative absence of archeological features which could be used to help
further rule out the possibility of this being the NT tombsite. Two further
points also bear noting here. The first is that on a priori grounds, the sisters
(Mariam and Salome, say) are perhaps less likely to occur in a NT tombsite
due to the possibility that they may have been married and hence been with
families of their own. The second is that if the disputed James ossuary were
to prove authentic, then James could no longer be an a priori candidate. (A related consideration arises if James was buried at the place of his execution.) Needless to say, if any of these out-of-sample names were "removed" from our a priori lists, or otherwise "downweighted," our "tail areas" would all decrease.

Let us next consider the impact of some of the assumptions. First, as concerns assumption A.8 (that Yoseh and Yehosef do not refer to the same person), the situation is somewhat subtle. While reasonable arguments may be advanced in favor of this assumption, if we were to choose to carry out an analysis without it, the probability structure under $H_0$ could then no longer be approximated by independence. Specifically, the drawings of the father and of the singletons would then become dependent in a way which cannot be specified in an obvious manner so that the combined RR value for Yoseh and the father Yehosef could then not be approximated by ordinary multiplication. One could, however, carry out analyses under two eventualities—the first (as we have done) under the assumption that these persons differ, and the second under the assumption that they are in fact the same. In the latter case, the father Yehosef in the generational ossuary would then become regarded as being the biblical brother (with only Yoseh, and not Yehosef, contributing toward the RR value), and the son Yeshua would then not count toward the RR value (or might count but in only a diminished way). Thus overall, without assumption A.8, the computations would not result in "significance."

Curiously, assumption A.4—regarding the Yehuda son of Yeshua ossuary—involves less computational complexity than at first seems since our analyses may in fact be carried out allowing for the presence of a full "generationally aligned" sequence "A son of B son of C." Because the NT genealogy has no *known* father-and-son pair with *both* dying between 30 CE and 70 CE, the youngest of this aligned trio—namely "A"—would *never* contribute toward the RR value. Hence the results of such analyses would actually be identical to those already carried out. A quite different conclusion would be reached, however, if the presence of this ossuary in the tomb was permitted to count "negatively," that is, in the direction of invalidating the find.

Concerning our specialized independence assumption A.9, a referee has argued that if the population of Jerusalem consisted of a small number of large clans, each sharing only a few ancestors, it could lead to name clustering, and the longitudinal dependences would then result in cross-sectional dependence as well. Of course, the cross-sectional approximate independence is ultimately a judgement call which we would have preferred to avoid, except that doing so would then limit the power of statistical procedures that can be devised. The data base for "assessing" this assumption more broadly (for the era in question) is limited, but it is not null. The series of "begats" in the NT are one potential data source which could be studied. More usefully, Ilan's (2002) compilation allows us to reconstruct some name matchings.

Thus, of the 23 entries of (generic) Yeshua derived from ossuaries, 13 are matched with the name of either a son or a father, with two of these being matched with both a father as well as a son. (One further entry is matched with a Salome, presumably a wife or sister.) From that data a slight tendency may be discerned for fathers called Yeshua to also name their sons Yeshua, but little else of significance is in evidence. Of the 45 entries of (generic) Yosef derived from ossuaries, 32 are matched with the name of either a son or a father, with one of these being matched with both a father and a son. [In two cases a daughter is mentioned (both times Martha). In another case a twin is mentioned (Eleazar), and in a related case two sons are mentioned (Eleazar and Joseph).] Two of these 32 cases indicate a son to be Yeshua (one corresponding to Talpiyot); none show Joseph as being a son of Yeshua. There appears to be a significant tendency for the sons and the fathers of (these ossuary-derived) Yosefs to have such rather unusual names as Shabi, Yoezer, Kallon, Agra, Benaiah, and so on. The impact of this on our analysis is conservative since the direction of the dependence implied only renders the Talpiyot observations more rare.

The last assumption we discuss here is A.7 concerning the name of Mary Magdalene. This assumption was suggested to us under the rationale outlined in Section 6 and it is the case that without the "rareness and relevance" of the Mariamenou [η] Mara inscription our test procedures would not prove "significant." Having no germane historical expertise, the author worked under this assumption, but the question may fairly be put as to whether or not it arose under the influence of the data. For inferences to be valid, the renditions for Mary Magdalene (particularly the most specialized ones) must, of course, be specified a priori. As this point will no doubt be argued by others it is unnecessary for us to belabour it here; however we offer two comments. First, our analysis does indeed assume the name of Mary Magdalene to have been either Mariamne or Mariamen (or a closely related rendition), a point legitimately subject to corroboration—or otherwise—by historical scholars. Should such scholarship ultimately prove inconclusive, an approach along the following lines may perhaps be considered: We have at our disposal a list of some 80 Mariams of the era whose actual name renditions are known to us; this includes the two Mariams from the Talpiyot find. If now we sought to categorize these 80 renditions according to the degree to which they appear to be appropriate ones for Mary Magdalene then it might well be that the rendition Mariamenou [η] Mara would be the one selected as being the most so. Here again, it would be the remarkable character of that rendition that would lead us to offer it that consideration. A separate issue is whether or not Mary Magdalene's candidature is legitimately a priori; while the logic behind the hypothesis APH 5 of Section 10 is "best efforts"-based, it is not incontestable.

The issues arising from the remaining assumptions, as well as their impacts on the analysis are more straightforward. We only remark, yet again, that all of the assumptions must be met for our "tail areas" to be meaningful.

Finally, concerning the (disputed) ossuary of James, it has been speculated that it might actually provenance to the Talpiyot site. On the basis of the currently available evidence the author does not believe any such claim to have been established, but its impact on the computations can nevertheless be described. First, with that ossuary included the statistical "significance" of the find would strengthen substantially even though the number of ossuaries conditioned upon would also have increased. No additional "RR" value would accrue for the common father, although some modest contribution might accrue on account of two patronymic ossuaries then likely being brothers. As for the (disputed) "brother of Jesus" component of the inscription, no further "RR" value would accrue from the repeated mention of Jesus. Of course, the mere mention of *that* particular name, *and in this way*, would obviously be considered to be sufficiently remarkable that any further statistical efforts would be rendered unnecessary.

Let us finally turn to the question of how one may interpret the "tail areas" computed in the preceding section, that is, the proportions ("under $H_0$") of obtaining "surprisingness" values as great as at Talpiyot. The issues here are not straightforward. Suppose, for the sake of this discussion, that agreement has been reached with respect to *all* of the hypotheses, assumptions, and conditions under which our computations were carried out; we shall hereafter collectively refer to these as our *provisos*. Using our "baseline" case for purposes of illustration, our computations suggest that a clustering of names as "surprising"—that is, "as relevant and as rare"—as those at Talpiyot occurs (approximately) once per 1,821,000 tombs under random sampling from the onomasticon. This number is considerably greater than the number of persons—let alone families—that died during the relevant span.[38]

We are, in fact, now in a position to carry out a particular *hypothesis test*: Here $H_0$ is the hypothesis that *all* 1,100 tombs in the vicinity of Jerusalem arose under random assignment of names, and $H_1$ is the hypothesis that *one unspecified one among these* 1,100 *tombs is that of the NT family*. The test statistic we shall use for this purpose is the lowest $H_0$-tail area for the RR values of the 1,100 tombs. A $p$-value for this test is *bounded above*[39] by the probability that one among these 1,100 tombs would have an RR value corresponding to an $H_0$-*tail area* less than or equal to 1/1,821,000; this probability bound is 1/1,655. *We therefore conclude, subject to the stated provisos, that there exists a NT tombsite, and furthermore that it is one of the* 1,100 *tombs in the vicinity of Jerusalem*. This is the first step in our inference, although it may be bypassed if we are prepared to accept the stated conclusion.

---

[38]Hence if, for example, the entire population could be divided into 10,000 Talpiyot-size tombs, the probability is 1/182 (under random assignment) that another family would have matched this tail area, and 1/1,655 that such a family would have occurred among the 1,100 existing tombs. Of course, larger families could have better odds that some *deliberately selected* subset of their names might be deemed to be as "surprising."

[39]It is *bounded above* because not all existing tombs have as yet been "measured," and one or more among them could conceivably provide a still lower tail areas. The fact

Interestingly—if counterintuitively—we cannot as an *immediate* next step conclude from this that the tomb at East Talpiyot must be that one. Our finding does, however, permit us to *objectively* assign a probability of 1/1,100 of being that of the NT family to any randomly selected one among these existing 1,100 tombs. Constructing a formal hypothesis test for whether or not the East Talpiyot tomb is actually that one is however not straight-forward[40]—a price we pay for the absence of a probability model (for RR values) under the "NT hypothesis." We are thus faced with the situation that we know (with $p = 1/1,655$) that one of the 1,100 tombs in the vicinity of Jerusalem is the NT family tombsite, and furthermore know that this knowledge was derived from an (extreme) RR tail area measurement which occurred *at a single tombsite*. And yet we cannot *immediately* conclude from this that this one tombsite must be that of the NT family. We do however know that the NT tombsite is either *the one* at East Talpiyot or *one of the others* among the 1,100 tombs in the vicinity of Jerusalem; unless a "type 1" error has occurred in our "first step," no other options are available.[41]

The second step in our inference involves the Bayes formula

$$\frac{P(A|B)}{P(\overline{A}|B)} = \frac{P(A)}{P(\overline{A})} \times \frac{P(B|A)}{P(B|\overline{A})}$$

for updating prior odds by a likelihood ratio. Here $A$ is the event that the Talpiyot tomb is that of the NT family, and $\overline{A}$ is the event that it is not. The conditioning event $B$ can be chosen in more than one way here. The

---

that not all tombs were configured identically complicates our arguments, however such conditioning is accepted statistical practice.

[40] There are analogies between our problem and one arising in "DNA matching" where a probability $P(A|B)$ is computed, although $P(B|A)$ is the one desired. In our application, what has been computed is the probability of obtaining an equally "surprising" cluster of names *given* that the tomb is not of the NT family while what is desired is the probability that this is the NT family tomb *given* that the cluster of names is so surprising. Some considerations that apply in such DNA studies therefore carry over to our problem. However our problem differs from the DNA one in that the DNA profile of the "accused party" is fully known, while the a priori profile for the NT tombsite is not.

[41] We shall not consider here the possibility that the foregoing arguments (as well as some others below) may be repeated using the 100 tombs already excavated in lieu of the 1,100 "in existence."

"natural" choice—where $B$ is the event of obtaining the specific cluster of names found at Talpiyot—is awkward to work with. We shall condition instead on the event that the $H_0$-tail area of the tomb being examined is less than or equal to that which occurred at Talpiyot. In proceeding, the following notation will be useful. Let $n_1$ be the number of tombs in the vicinity of Jerusalem that have already been excavated; that number[42] is approximately 100. Let $n_2$ be the number of tombs—approximately 1,100—that exist in the vicinity of Jerusalem. Let $n_3$ be the number of tombs (of "Talpiyot size") that could have been formed had the *entire* population of Jewish adults been buried in tombs with inscribed ossuaries; that number is somewhat less than 10,000. Let $q$ be the $H_0$-tail area of the RR statistic for the Talpiyot tomb according our baseline, or to any other "case" being considered; the order of magnitude of $q$ is about $10^{-6}$. In this notation, the $p$-value for our test at step one is $p = n_2 q$, while our odds-updating formula becomes

$$\frac{P(A|B)}{P(\overline{A}|B)} = \frac{1}{(n_2 - 1)} \times \frac{\theta}{q} = \frac{\theta}{(n_2 - 1)q},$$

where $\theta \equiv P(B|A)$ is the probability that a NT family tomb would consist of a cluster of names as surprising (based on our RR approach) as that at Talpiyot. Some readers may believe that $\theta = 1$, or in that order of magnitude; for them the inference process will now be completed. A similar remark applies to readers prepared to at least believe that $\theta$ is not terribly small.

Readers who prefer not to assume that $\theta$ is not very small may consider, as a third step, to obtain a lower confidence bound for $\theta$. Among the $n_2$ existing tombs, that of the NT family has probability $\theta$ of "attaining $q$" while the probability that one among the $n_2 - 1$ others does is given by $(n_2 - 1)q$ since their tail areas are uniformly distributed. Hence the probability that the tail area value of $q$ will be attained in the group of all $n_2$ existing tombs is given by

$$\tau \equiv \theta + (n_2 - 1)q - (n_2 - 1)q\theta = \theta[1 - (n_2 - 1)q] + (n_2 - 1)q.$$

This in fact is the probability of a Bernoulli event. A decidedly conservative $100(1 - \alpha)\%$ lower confidence bound for $\tau$ is given by 0 if the "$q$-event" is not attained, and by $\alpha$ if (as in our case) it is. Solving $\tau \geq \alpha$ then gives the $100(1 - \alpha)\%$ lower confidence bound

$$\theta \geq \frac{\alpha}{1 - (n_2 - 1)q} - \frac{(n_2 - 1)q}{1 - (n_2 - 1)q}$$

---

[42] The ossuary-sourced listings in Ilan also divide up into approximately 100 Talpiyot-like configurations.

for $\theta$, from which we obtain the confidence bound

$$\frac{P(A|B)}{P(\overline{A}|B)} \geq \frac{\alpha - \beta}{\beta(1-\beta)}, \qquad \text{where } \beta \equiv (n_2 - 1)q,$$

for the odds ratio; for small $\beta$, this bound is approximately $(\alpha/\beta) - 1$. For illustration, in our baseline case, $n_2 = 1,100$, and $q = 1/1,821,000$; if $\alpha = 0.05$ or $0.01$, the lower confidence bound for $\theta$ is $0.0494$ or $0.0094$, and in turn the lower confidence bound for $P(A|B)/P(\overline{A}|B)$ will be $81.90$ or $15.58$, respectively. If we had assumed instead that $\theta = 1$, $0.5$, or $0.1$, then using the value $\theta/\beta$ we would have obtained odds ratios of $1657$, $828$ and $167$, respectively. These results are, of course, all dependent upon our provisos.

To summarize now, in this paper we have conveyed an interesting data set and have provided some background essential for its interpretation. We have also proposed a paradigm intended to deal with the purely statistical questions such data pose—that based on "surprisingness," or the "RR" (relevance and rareness) measure. Although related to classical methods, this paradigm differs from them in a number of ways. In practice, there are probably few real-data-based analyses of consequence on controversial issues which do not lend themselves to counterargumentation. The results of our analysis could be challenged on the basis of the methodology applied or the assumptions on which it was based. We hope that the statistical methodology itself will not be found unduly controversial. As concerns the assumptions, the situation is different; while we have provided a rationale for each, they are not unassailable. Furthermore, arguments could be mounted to the effect that no a priori lists of persons and name renditions could ever be legitimately assembled after the fact. The influence of the Mariamenou [$\eta$] Mara inscription in the analysis particularly flags it as a "target."

If the assumptions A.1–A.9 under which our computations have been carried out are accepted, and if an a priori list of NT tomb candidates, together with an a priori set of name renditions for them were accepted as well, and further, if the list of candidates contained at least those key persons which the Talpiyot inscriptions seemingly allude to, then our computations strongly suggest that the possibility that the Talpiyot tomb is that of the NT family merits serious consideration. Subject to the stated provisos, our numerical experiments also suggest that this conclusion is robust to *moderate* variations in the specifications of the lists of candidates and name rendition categories. It is also reasonably robust with respect to variations in the relative frequencies for these name renditions and with respect to "reasonable" variations in the components of our definition of "surprise" (or "RR" value).

Even if statistical significance of the "RR" value of the Talpiyot tomb were accepted as fact, nothing in the purely statistical aspects of our analysis *directly* addresses such questions as whether or not Jesus and Mary Magdalene might have been married, or whether or not they may have had a son; certainly other possible explanations exist as well. Further, statistical significance only establishes that either the null hypothesis must be false, or we have observed an event of rare chance; either of these are possibilities.

Among the various assumptions made, perhaps the one that most "drives" our analysis in the direction of "significance" is the extraordinary inscription Mariamenou [$\eta$] Mara. It has been speculated that Mary Magdalene was a principal driving force in the movement founded by Jesus but was later vilified in the course of patriarchal power struggles. While we are in no position to weigh in on any such theories, what we can say is that from a purely statistical point of view, this much is true: It is the presence in this burial cave of the ossuary of Mariamenou [$\eta$] Mara, and the mysteries concerning the identity of the woman known as Mary Magdalene, that hold the key for the degree to which statistical analysis will ultimately play a substantive role in determining whether or not the burial cave at East Talpiyot happens to be that of the family of Jesus of Nazareth.

**Acknowledgments.**  For helpful discussions and other assistance the author is indebted to David Andrews, Nicole Austin, James Charlesworth, Radu Craiu, Tom DiCiccio, Laurel Duquette, Grace Feuerverger, Steve Fienberg, Camil Fuchs, Don Fraser, Shimon Gibson, Felix Golubev, Itay Heled, Tyler Howard, Tal Ilan, Simcha Jacobovici, Marek Kanter, Georges Monette, Hadas Moshonov, Radford Neal, Nancy Reid, Ben Reiser, James Tabor, Robert Tibshirani, the staff at Associated Producers Ltd., the editors of this Journal and the referees, and certain others who prefer to remain anonymous. Due to confidentiality agreements the author was obliged to respect, not all those named were aware of the nature of this work; it goes without saying that the author alone is responsible for the contents of this paper. I particularly wish to reiterate my indebtedness to Simcha Jacobovici for bringing this extraordinary data set to my attention, for sharing his extensive knowledge base regarding this archeological find, and for facilitating scholarly contacts; the assumptions under which our analysis was carried out were proposed by him. I also thank Associated Producers for permission to reproduce the images of the six inscribed ossuaries. Special thanks to Tyler Howard for his astute suggestions and highly conscientious work in checking my original S-PLUS code and converting it into R code. I also take this opportunity to reaffirm my indebtedness to David Andrews for the many incisive insights he has generously shared with me over the years on the subtleties of statistically significant applications; no finer statistician have I ever known. Likewise, S. Fienberg and G. Monette were most gracious in supplying invaluable comments and suggestions on an earlier draft. Last, but not least, I wish to thank Professor James Tabor for having so generously shared with me his wealth of knowledge concerning historical matters of the New Testament era, and for tirelessly responding to many queries, particularly during the process in which the a priori hypotheses of Section 10 were being formulated.

## SUPPLEMENTARY MATERIAL

**Computing code for "Statistical analysis of an archeological find"** (doi: 10.1214/08-AOAS99supp; .txt). This file contains the R computing code used to produce the results in this paper. The code is self-explanatory and is easily modified to generate the reported results. It may also be modified to account for different assumption sets to enter into the "RR" (relevance and rareness) computations.

DEPARTMENT OF STATISTICS
RM. 6009, SIDNEY SMITH BLDG.
100 ST. GEORGE STREET
UNIVERSITY OF TORONTO
TORONTO, ONTARIO
CANADA M5S 3G3
E-MAIL: andrey@utstat.toronto.edu